\documentclass[reprint,aps,prd,amsmath,amssymb,floatfix,nofootinbib,preprintnumbers]{revtex4-2}

\usepackage{graphicx}
\graphicspath{ {./img/} }
\usepackage[dvipsnames]{xcolor}
\usepackage{hyperref}
\hypersetup{colorlinks, urlcolor=BlueViolet, citecolor=Plum, linkcolor=PineGreen}

\usepackage{bm}
\usepackage{cleveref}

\newcommand{\rep}[2]{(\mathbf{#1}, \mathbf{#2})}
\newcommand{\RvRd}{(\mathbf{R}_v, \mathbf{R}_d)}

\begin{document}

\title{Explaining the cosmological dark matter coincidence in asymmetric dark QCD}

\author{Alexander C.~Ritter}
\email{ritter@unimelb.edu.au}
\affiliation{ARC Centre of Excellence for Dark Matter Particle Physics, School of Physics, The University of Melbourne, Victoria 3010, Australia}

\author{Raymond R.~Volkas}
\email{raymondv@unimelb.edu.au}
\affiliation{ARC Centre of Excellence for Dark Matter Particle Physics, School of Physics, The University of Melbourne, Victoria 3010, Australia}

\begin{abstract}
    To properly solve the coincidence problem ($\Omega_\mathrm{DM} \simeq 5\Omega_\mathrm{VM}$) in a model of asymmetric dark matter, one cannot simply relate the number densities of visible and dark matter without also relating their particle masses. Following previous work, we consider a framework where the dark matter is a confined state of a dark QCD gauge group whose confinement scale is dynamically related to the QCD confinement scale by a mechanism utilising infrared fixed points of the two gauge couplings. In this work we present a new, `zero-coupling infrared fixed point' approach, which allows a larger proportion of models in this framework to generically relate the masses of the visible and dark matter particles. Due to the heavy mass scale required for the new field content, we introduce supersymmetry to the theory. We consider how these models may be incorporated in a full theory of asymmetric dark matter, presenting some example leptogenesis-like models. We also discuss the phenomenology of these models; in particular, there are gravitational wave signals which, while weak, may be measurable at future mHz and $\mu$Hz detectors.
\end{abstract}

\maketitle


\section{Introduction}

The nature of dark matter (DM) remains unknown. While its existence has been firmly established by a range of astrophysical observations, these depend only on its gravitational interactions, and -- as the conventional wisdom asserts -- provide no indication as to its non-gravitational interactions other than their weakness. However, one clue to its identity that may exist in the present cosmological data is the similarity between the present-day mass fractions of visible matter (VM) and dark matter~\cite{Planck:2018vyg},
\begin{equation}\label{eqn:cosmo_coincidence}
    \Omega_\mathrm{DM} \simeq 5\Omega_\mathrm{VM},
\end{equation}
which is suggestive of a deep connection between the origins of VM and DM. We refer to this fact as the \emph{cosmological coincidence}; indeed, this $\mathcal{O}(1)$ ratio arises coincidentally for most canonical dark matter candidates, such as the WIMP or axion, where the dynamics that generate the cosmological abundances of visible and dark matter are entirely unrelated to each other.

The predominant paradigm that attempts to motivate this relationship is asymmetric dark matter (ADM)~\cite{Petraki:2013wwa, Zurek:2013wia}, where the DM and VM abundances both result from a particle--anti-particle asymmetry generated by some baryogenesis-like mechanism. In ADM models, it is ensured that the number densities are similar, $n_\mathrm{DM} \sim n_\mathrm{VM}$, either dynamically or through the imposition of a symmetry. However, such a mechanism alone does not resolve the cosmological coincidence; to explain the relationship of Eqn.~\ref{eqn:cosmo_coincidence} also requires an explanation for $m_\mathrm{DM} \sim m_\mathrm{VM}$\footnote{There have also been attempts to resolve the coincidence problem that directly relate the cosmological mass densities of VM and DM without requiring the number densities and particle masses to both be of a similar order. These arise from anthropic arguments \cite{Linde:1987bx, Wilczek:2004cr, Hellerman:2005yi} or from some novel dynamics \cite{Rosa:2022sym, Brzeminski:2023wza}.}.

The main approach to this problem is to introduce a dark QCD-like gauge group $SU(N_d)$ and some mechanism that relates the confinement scale $\Lambda_\mathrm{dQCD}$ to the QCD confinement scale $\Lambda_\mathrm{QCD}$; the DM candidate in these models is thus a dark baryon whose mass is similar to the proton mass~\cite{Bai:2013xga, Newstead:2014jva, Ritter:2022opo, Hodges:1993yb, Berezhiani:1995am, Mohapatra:2000rk, Berezhiani:2000gw, Ignatiev:2003js, Foot:2003jt, Foot:2004pq, Foot:2004pa, An:2009vq, Cui:2011wk, Lonsdale:2014wwa, Lonsdale:2014yua, Lonsdale:2017mzg, Lonsdale:2018xwd, Ibe:2018juk, Ibe:2019ena, Ritter:2021hgu, Ibe:2021gil, Bodas:2024idn, Farina:2015uea, Garcia:2015toa, Farina:2016ndq, Terning:2019hgj, Beauchesne:2020mih, Feng:2020urb, Bittar:2023kdl, Alonso-Alvarez:2023bat, Murgui:2021eqf}. The related confinement scales in these models usually derive from a $\mathbb{Z}_2$ symmetry between the visible and dark QCD gauge groups, often in the context of mirror matter\footnote{In mirror matter models, the dark sector consists of a copy of the SM related to the SM by a $\mathbb{Z}_2$ `mirror' symmetry. Such a structure is primarily considered \textit{a priori}, or to reintroduce a fundamental parity symmetry~\cite{Hodges:1993yb, Berezhiani:1995am, Mohapatra:2000rk, Berezhiani:2000gw, Ignatiev:2003js, Foot:2003jt, Foot:2004pq, Foot:2004pa, An:2009vq, Cui:2011wk, Lonsdale:2014wwa, Lonsdale:2014yua, Lonsdale:2017mzg, Lonsdale:2018xwd, Ibe:2018juk, Ibe:2019ena, Ritter:2021hgu, Ibe:2021gil, Bodas:2024idn}, but it can also be used to address the little hierarchy problem in the context of Twin Higgs models~\cite{Farina:2015uea, Garcia:2015toa, Farina:2016ndq, Terning:2019hgj, Beauchesne:2020mih, Feng:2020urb, Bittar:2023kdl, Alonso-Alvarez:2023bat}.}, or from some unification at high energies~\cite{Lonsdale:2014wwa, Lonsdale:2014yua, Bodas:2024idn, Murgui:2021eqf}. However, avoiding the imposition of a high-energy symmetry, there is a dynamical approach, introduced by Bai and Schwaller~\cite{Bai:2013xga}, that takes advantage of the possible existence of infrared fixed points (IRFPs) in QCD-like theories~\cite{Bai:2013xga, Newstead:2014jva, Ritter:2022opo}.

In the framework of Ref.~\cite{Bai:2013xga}, new field content is introduced, charged under one or both of visible and dark QCD. This leads to the coupled renormalisation group (RG) evolution of the coupling constants of each gauge group towards a fixed point in the IR, thereby relating the coupling values, and thus the confinement scales, at low energies. One of the appeals of this approach is its insensitivity to the initial coupling strengths of the two gauge groups in the UV, and it is assumed in Ref.~\cite{Bai:2013xga} that the results in the IR are completely independent of the initial conditions at high energies. However, in our first paper~\cite{Ritter:2022opo} we showed that this is not the case. After accounting for the dependence of the results on the initial UV gauge couplings, we reassessed the ability of models in this framework to naturally relate $\Lambda_\mathrm{dQCD}$ and $\Lambda_\mathrm{QCD}$. Considering a simplified version of the framework with $N_d = 3$ and only new field content in fundamental representations of visible and dark QCD, we found a set of `viable' models where one would reasonably expect confinement scales on a similar order of magnitude.

However, our results were not entirely satisfactory. Firstly, the viable models comprised a very small fraction of the models in the framework, and to avoid stringent collider constraints on the new coloured fields, some of the new fields required very high multiplicities. Together, these issues made these models unwieldy and difficult to build upon further. This is a problem, as these models only relate the particle masses of VM and DM; to serve as a complete explanation of the cosmological coincidence problem, we must embed these models within an extended ADM theory.

In this work, our aim is to find models that may be more amenable to further study. We relax the restrictions on the number of dark colours $N_d$ and the representations of the new field content, and consider the ADM mechanisms that could be incorporated in these models. In the pursuit of a framework that more generically produces viable models, we are led to investigate models where both couplings are zero at the IRFP; this represents a new framework, beyond that explored by Bai and Schwaller. We refer to this as the zero-coupling fixed point (ZCFP) approach, distinguishing it from the finite-coupling fixed point (FCFP) approach which we have previously explored. We find that the ZCFP approach can alleviate some of the issues associated with the FCFP approach, notably that (i) viable models can be found more generically, and only require a few additional fields to be introduced, and (ii) the mass scale of new physics is generally greater than a TeV, which prevents the phenomenological tension that was difficult to avoid in the FCFP approach. 

However, this second point introduces its own difficulties, as the required mass scale for new physics is generically very high, which introduces a naturalness issue of its own in the running of the Higgs mass parameter. We first look for models where the mass of the new field content is low enough to avoid these issues -- that is, TeV-scale -- and have some success. However, while these models do exist, they are rare, and so do not improve on the first issue associated with the FCFP approach mentioned above. To consider models with very heavy new field content we then introduce supersymmetry to the framework. Here, we find some quite promising results; in particular, when $N_d = 4$ we find a fairly generic set of viable models.

The paper is organised as follows: in Section~\ref{sec:framework} we describe the framework, before evaluating the results of our previous work, and their shortcomings, in Section~\ref{sec:finite_coupling}. In Section~\ref{sec:zero_coupling} we introduce and discuss the zero-coupling fixed point approach, and then search for and identify viable ZCFP models in Section~\ref{sec:models}. Following this, in Section~\ref{sec:adm} we consider how to provide these viable models with an ADM mechanism, focusing on leptogenesis-like mechanisms similar to Ref.~\cite{Bai:2013xga} in both SUSY and non-SUSY cases. We finish by briefly highlighting some of the phenomenology of this framework, including possible direct-detection and gravitational wave signals, in Section~\ref{sec:pheno}, before concluding in Section~\ref{sec:conclusions}.


\section{The framework}\label{sec:framework}

The basic ingredients of the framework are a new dark confining gauge group $SU(N_d)$ which we refer to as `dark QCD', and a selection of field content charged under one or both of visible and dark QCD. The field content consists of Dirac fermions and complex scalars. For simplicity, we take all the new fields to share a (heavy) mass $M$, bar a number of Dirac fermions ($n_{dq}$) charged only under the fundamental representation of dark QCD. We take these latter fields to be `light' in the sense that their mass is lower than the dark confinement scale $\Lambda_\mathrm{dQCD}$. We refer to these fields as `dark quarks' and they confine into a dark baryon with a mass on the scale of $\Lambda_\mathrm{dQCD}$, which serves as the dark matter candidate.

We denote a `model' within this framework by its selection of field content. The purpose of these models is to naturally obtain similar confinement scales for visible and dark QCD; the dark baryon then has a mass that is on a similar order of magnitude to the proton mass. We denote a model that achieves this goal a `viable' model.

In this framework, the similarity of the confinement scales results from a dynamical mechanism that relates the coupling constants of each gauge group at low energies. The key equations are the $\beta$-functions that determine the RG evolution of the couplings. In our previous work~\cite{Ritter:2022opo}, we specialised to $N_d = 3$ and only considered field content in the fundamental representations of visible and dark QCD; in this work, we relax those assumptions. Here we give the $\beta$-functions for the strong coupling constant $\alpha_s$ and dark coupling constant $\alpha_d$, $\beta_v(\alpha_s, \alpha_d) \equiv d\alpha_s/d\log\mu$ and $\beta_d(\alpha_s, \alpha_d) \equiv d\alpha_d/d\log\mu$ respectively, which are coupled at two-loop level due to the presence of fields charged under both gauge groups. These are presented for an arbitrary selection of field content of arbitrary representation, where $n_{f/s}^{\RvRd}$ is the multiplicity of a Dirac fermion ($f$) or complex scalar ($s$) in the $\RvRd$ representation of $G_v \times G_d = SU(3)_\mathrm{QCD}\times SU(N_d)_\mathrm{dQCD}$\footnote{In this work, some of the representations we consider are real, like the $\rep{1}{6}$ when $N_d = 4$. Fields in such representations could be included as Majorana fermions or real scalars. We do not account for models with such fields in this framework. As we shall see in Section~\ref{sec:adm}, in realistic models accommodating a baryogenesis mechanism the fields we introduce generally have a non-zero hypercharge, and so are not in real representations of the full gauge theory.}, and $\{n_{f/s}\}$ is the set of all multiplicities for the fermions or scalars included in the model. The two-loop $\beta$-function for $\alpha_s$ is then given by~\cite{Jones:1981we}
\begin{widetext}
\begin{align}\label{eqn:full_beta}
    \beta_v(\alpha_s, \alpha_d) &= \frac{\alpha_s^2}{2\pi}\Bigg[\sum_{\{n_f\}}n_f^{\RvRd}\frac{2}{3}T(\mathbf{R}_v)2d(\mathbf{R}_d) + \sum_{\{n_s\}}n_s^{\RvRd}\frac{1}{3}T(\mathbf{R}_v)d(\mathbf{R}_d) - \frac{11}{3}C_2(G_v)\Bigg]\nonumber\\
    &+ \frac{\alpha_s^3}{8\pi^2}\Bigg[\sum_{\{n_f\}}n_f^{\RvRd}\left(\frac{10}{3}C_2(G_v) + 2C_2(\mathbf{R}_v)\right)T(\mathbf{R}_v)2d(\mathbf{R}_d)\nonumber\\
    &\hspace{2cm}+ \sum_{\{n_s\}}n_s^{\RvRd}\left(\frac{2}{3}C_2(G_v) + 4C_2(\mathbf{R}_v)\right)T(\mathbf{R}_v)d(\mathbf{R}_d) - \frac{34}{3}C_2(G_v)^2\Bigg]\nonumber\\
    &+ \frac{\alpha_s^2\alpha_d}{8\pi^2}\Bigg[\sum_{\{n_f\}}n_f^{\RvRd}2C_2(\mathbf{R}_d)T(\mathbf{R}_v)2d(\mathbf{R}_d) + \sum_{\{n_s\}}n_s^{\RvRd}4C_2(\mathbf{R}_d)T(\mathbf{R}_v)d(\mathbf{R}_d)\Bigg],
\end{align}
\end{widetext}
and $\beta_d(\alpha_s, \alpha_d)$ is found by exchanging $\alpha_s$ and $\alpha_d$ and interchanging the indices $v \leftrightarrow d$. Here, $C_2(G_v) = N_c = 3$ and $C_2(G_d) = N_d$ are the quadratic Casimirs of the adjoint representations, and for a given representation $\mathbf{R}$, $d(\mathbf{R})$ is the dimension of the representation, $C_2(\mathbf{R})$ is its quadratic Casimir, and $T(\mathbf{R})$ is its Dynkin index.

In this work, we also consider supersymmetric versions of the framework; this avoids the naturalness issue that arises in models with a very high mass scale $M$. In this case, the new field content consists of chiral supermultiplets, each containing a chiral fermion and a complex scalar. We specify the multiplicity of a chiral supermultiplet in the $\RvRd$ representation of $SU(3)_\mathrm{QCD}\times SU(N_d)_\mathrm{dQCD}$ by $n^{\RvRd}$, and the set of all multiplicities of the supermultiplets included in the model by $\{n\}$. The two-loop $\beta$-function for $\alpha_s$ is then given by~\cite{Jones:1981we}
\begin{widetext}
\begin{align}\label{eqn:full_susy_beta}
    \beta_v(\alpha_s, \alpha_d) &= \frac{\alpha_s^2}{2\pi}\Bigg[\sum_{\{n\}}n^{\RvRd}T(\mathbf{R}_v)d(\mathbf{R}_d) - 3C_2(G_v)\Bigg]\nonumber\\
    &+ \frac{\alpha_s^3}{8\pi^2}\Bigg[\sum_{\{n\}}n^{\RvRd}\left(2C_2(G_v) + 4C_2(\mathbf{R}_v)\right)T(\mathbf{R}_v)d(\mathbf{R}_d) - 6C_2(G_v)^2\Bigg]\nonumber\\
    &+ \frac{\alpha_s^2\alpha_d}{8\pi^2}\Bigg[\sum_{\{n\}}n^{\RvRd}4C_2(\mathbf{R}_d)T(\mathbf{R}_v)d(\mathbf{R}_d)\Bigg],
\end{align}
\end{widetext}
and $\beta_d(\alpha_s, \alpha_d)$ is found by exchanging $\alpha_s$ and $\alpha_d$, and interchanging the indices $v \leftrightarrow d$, as before.

The mechanism by which these models naturally relate the confinement scales was first introduced by Bai and Schwaller~\cite{Bai:2013xga}. The general idea is that, for certain models, the selection of field content is such that the coupled $\beta$-functions exhibit an infrared fixed point (IRFP). The couplings, which may be be initially unrelated in the UV, evolve towards the IRFP, and are thus dynamically related to each other in the IR, despite the absence of a discrete symmetry that interchanges the visible and dark QCD gauge sectors. After the new field content decouples below its mass scale $M$, the IRFP is broken and the gauge couplings run until they each become non-perturbative and their associated gauge interactions become confining; the relationship between the couplings at the IRFP then leads to a relationship between the confinement scales.

In their initial work, Bai and Schwaller assumed that the couplings reach their IRFP values by the decoupling scale, which was generally TeV-scale. However, in Ref.~\cite{Ritter:2022opo} we showed this to be untrue; the couplings only approach their fixed point values without reaching them. The dark confinement scale predicted for a given model then depends upon the initial conditions for the couplings in the UV. Accounting for this dependence, we then considered a viable model to be one that obtains similar confinement scales for the two sectors for a decent proportion of the $(\alpha_s^\mathrm{UV}, \alpha_d^\mathrm{UV})$ parameter space. In this context, we defined a heuristic to judge the viability of a given model which we called the `viability fraction' $\epsilon_v$: it is defined as the proportion of the UV coupling parameter space for which $\Lambda_\mathrm{dQCD}$ is between 0.2 and 5\,GeV.

The viability fraction $\epsilon_v$ is the main tool we use to assess the ability of given model to serve as an explanation of the cosmological coincidence problem. The idea is that in a model with a decently large value of $\epsilon_v$, there is a reasonable chance that the confinement scales will be on the same order of magnitude for a random selection of initial values for the coupling constants in the UV. Of course, the distribution from which these values are selected is unknown. The viability fraction we define is merely a simple implementation of this concept, but it is by no means the only one. We also note that these models do not solve the cosmological coincidence problem alone; rather, they provide a template that must be embedded within a full ADM model. In the full theory, a baryogenesis mechanism is required that generates related number densities for visible and dark matter. We return to this question in Section~\ref{sec:adm}.


\section{The finite-coupling infrared fixed point approach}\label{sec:finite_coupling}

In this section we provide an evaluation of the key conclusions of our previous work~\cite{Ritter:2022opo}. Our goal there was to identify models within the framework that could viably explain the coincidence problem. Here we identify common features of these viable models, and in particular focus on their shortcomings, namely:
\begin{enumerate}
    \item These viable models comprise a very small proportion of the total space of models.
    \item Most viable models require a sub-TeV mass scale for the new physics, i.e.\ new sub-TeV coloured states that would be copiously produced at colliders. While we do find viable models with $M >$~TeV, they generally require large multiplicities for the new fields introduced.
\end{enumerate}

Both of these issues make it difficult for these models to be incorporated within a full ADM model. The purpose of this section is to motivate the approach we take in the following section, which seeks to alleviate these issues.

\subsection{Obtaining a finite-coupling fixed point}

A fixed point of the coupled RG evolution of the gauge couplings of visible and dark QCD, denoted $(\alpha_s^\star, \alpha_d^\star)$, is given by
\begin{equation}\label{eqn:irfp_beta_condition}
    \beta_v(\alpha_s^\star, \alpha_d^\star) = \beta_d(\alpha_s^\star, \alpha_d^\star) = 0.
\end{equation}

Rewriting Eqn.~\ref{eqn:full_beta} as
\begin{equation}
    \beta_v(\alpha_s, \alpha_d) = b_v^{(0)}\frac{\alpha_s^2}{2\pi} + b_v^{(1)}\frac{\alpha_s^3}{8\pi^2} + b_v^{(1')}\frac{\alpha_s^2\alpha_d}{8\pi^2},
\end{equation}
and with $\beta_d(\alpha_s, \alpha_d)$ given by exchanging $\alpha_s$ and $\alpha_d$ and interchanging the indices $v \leftrightarrow d$, there are four solutions to Eqn.~\ref{eqn:irfp_beta_condition}:
\begin{align}
    \alpha_s^\star = 0,&\quad \alpha_d^\star = 0\label{eqn:zcfp_solution} \\
    \alpha_s^\star = -4\pi\frac{b_v^{(0)}}{b_v^{(1)}},&\quad \alpha_d^\star = 0\\
    \alpha_s^\star = 0,&\quad \alpha_d^\star = -4\pi\frac{b_d^{(0)}}{b_d^{(1)}}\\
    \alpha_s^\star = 4\pi\frac{b_v^{(0)}b_d^{(1)}-b_v^{(1')}b_d^{(0)}}{b_v^{(1')}b_d^{(1')}-b_v^{(1)}b_d^{(1)}},& \quad \alpha_d^\star = 4\pi\frac{b_d^{(0)}b_v^{(1)}-b_d^{(1')}b_v^{(0)}}{b_v^{(1')}b_d^{(1')}-b_v^{(1)}b_d^{(1)}}.\label{eqn:fcfp_solution}
\end{align}

Note that these are not necessarily infrared fixed points; they may be UV fixed points, or they may be saddle points of the coupled evolution of the gauge couplings. Indeed, these nominal solutions are not necessarily physical, let alone perturbative.

In our previous work, we considered models in which the last of these solutions -- Eqn.~\ref{eqn:fcfp_solution} -- is an infrared fixed point, with both $\alpha_s^\star$ and $\alpha_d^\star$ taking perturbative values\footnote{In this work, we consider gauge couplings $\alpha_s, \alpha_d < 0.3$ to be perturbative. This ensures the pertubative validity of our two-loop $\beta$-function calculations: see Sec.~V\,A in our previous work~\cite{Ritter:2022opo}.}. This is not the only case one could consider, and so we refer to this case specifically as the `finite-coupling fixed point' approach. This contrasts it with the case we consider in the next section, the `zero-coupling fixed point' approach, where the first of these solutions -- Eqn.~\ref{eqn:zcfp_solution} -- is an infrared fixed point.

In Eqn.~\ref{eqn:fcfp_solution} we immediately see one of the shortcomings that this approach faces; that is, the values of $\alpha_s^\star$ and $\alpha_d^\star$ have a highly non-trivial dependence on the various $\beta$-function coefficients. Firstly, this prevents any simple links being drawn between the field content of a given model and the value of the couplings at the fixed point, especially since a given field contributes to multiple $\beta$-function coefficients. Secondly, as stated already, for a generic model these values may not be perturbative, or even physical. To obtain a model that has a perturbative FCFP requires a delicate interplay between the $\beta$-function coefficients, and we find that this occurs in only a small fraction of models. For example, in our previous work we initially analysed a set of 12,288 models; of these models, 1354 ($\sim$10\%) have an infrared fixed point, and only 155 ($\sim$1\%) of the models have a perturbative fixed point.

The last observation we make here is that for the FCFP of Eqn.~\ref{eqn:fcfp_solution} to be an infrared fixed point, $b_v^{(0)}$ and $b_d^{(0)}$ must both be negative. The argument is as follows: if the theory contains an infrared FCFP, then initially small couplings $\alpha_s^\mathrm{UV}, \alpha_d^\mathrm{UV} \ll 1$ must increase towards the IRFP with decreasing energy scale (that is, the zero-coupling fixed point of Eqn.~\ref{eqn:zcfp_solution} must be a UV fixed point). So, for these small couplings, the $\beta$-functions must be negative; also, for small couplings the $\beta$-functions are dominated by the one-loop terms. Thus, the one-loop coefficients $b_v^{(0)}$ and $b_d^{(0)}$ must be negative.

\subsection{Large viability fractions}

We search for models with the largest possible value for the viability fraction. In our previous work, the largest value we found for $\epsilon_v$ was around 0.4, and we considered models with $\epsilon_v > 0.3$ to be viable. These comprised a smaller fraction again of the total models. For example, in the set of 12,288 models mentioned previously, of the 155 models with perturbative infrared FCFPs, only 19 had $\epsilon_v > 0.3$, comprising $\sim$0.1\% of the total models.

This restriction is due to the fact that the models with the largest viability fractions all have `small' IRFPs with $\alpha_s^\star, \alpha_d^\star \lesssim 0.1$. The reason for this is in two parts. Firstly, at the decoupling scale of the new physics, we need $\alpha_s < \alpha_s(M_Z) = 0.11729$ in order to be able to match the strong coupling to its measured value at the $Z$ mass, and thus to calculate $\Lambda_\mathrm{dQCD}$. So, for this to be achieved for a wide range of initial coupling values in the UV, $\alpha_s$ must be evolving towards a fixed point value $\alpha_s^\star < 0.11729$. Secondly, to have $\Lambda_\mathrm{dQCD} \sim \Lambda_\mathrm{QCD}$ for a large area of the UV coupling parameter space, we need $\alpha_s$ and $\alpha_d$ to be similar in size at the decoupling scale. This indicates that the couplings should be evolving towards fixed points with similar values. Since $\alpha_s^\star$ must be small, $\alpha_d^\star$ must also be small.

We note that this second point assumes that similar gauge couplings at the decoupling scale imply similar confinement scales in the two sectors. This is only true if the gauge coupling evolution is similar in the two sectors after decoupling. If $N_d \neq 3$, or if the number of light dark quarks differs substantially from the number of light SM quarks, then this is not necessarily true. Indeed, we find in general that the value of $\epsilon_v$ depends strongly on the number of light dark quarks; this will be discussed further in Section~\ref{sec:models}.

We also note that the intuitive concept of the success of a viable model now differs from the original idea of Bai and Schwaller. The relationship between the couplings in the infrared results from their evolution towards some fixed point whose couplings take small values; that is, they are `focused' toward a point near the origin of the $(\alpha_s, \alpha_d)$ coupling plane, and thus brought close to each other in value. This motivates us to consider the zero-coupling fixed point approach that we introduce in the next section, where the origin itself, $\alpha_s = \alpha_d = 0$, serves as the infrared fixed point that draws the couplings together from their values in the UV.

\subsection{The mass scale of new physics}

An issue that arises generically for these viable models is that in much of the $(\alpha_s^\mathrm{UV}, \alpha_d^\mathrm{UV})$ parameter space, the required mass scale of new physics $M$ is below a TeV. These new coloured particles with $\mathcal{O}(100)$\,GeV masses would be copiously produced at colliders, leading to the exclusion of these models. In our previous work, we thus searched for models in which $M$ is TeV-scale or higher.

The general idea is that to obtain a high mass scale, the running of the gauge couplings in the full theory must be `fast', so that $\alpha_s$ evolving from the UV matches to the required coupling to generate $\alpha_s(M_Z)$ at a sufficiently high scale. Strong running requires large values for the $\beta$-function coefficients. In the FCFP approach, the one-loop $\beta$-function coefficients $b_v^{(0)}$ and $b_d^{(0)}$ must be negative, but there is no restriction on the sign of the two-loop terms. So, the strongest running occurs when $b_v^{(0)}$ and $b_d^{(0)}$ are negative and close to zero. This, however, requires a fine cancellation between the gauge and matter contributions to these coefficients, and so again only occurs for select sets of field content, further reducing the proportion of total models that are viable and phenomenologically allowed.

In our first paper, we achieved strong running by considering models with high multiplicities for the jointly-charged $\rep{3}{3}$ scalars, while finding that successful models also had large multiplicities for the other heavy new scalars. Each additional $\rep{3}{3}$ scalar contributes to all $\beta$-function coefficients, and so strengthens the running more than scalars charged under only one gauge group.

We note that this means that we are considering models in which the one-loop coefficients are much smaller in magnitude than the two-loop coefficients, thanks to the precise cancellation between the gauge and matter contributions to the one-loop coefficient. This could lead to questions about the perturbative validity of this analysis. Our answer is to consider the magnitude of each term that contributes to a given coefficient, not their sum (as this ignores the artificial smallness that occurs due to cancellations between terms), and to ensure that individual two-loop terms are smaller in magnitude than individual one-loop terms.

The issue now is that not only do these successful models comprise a very small proportion of the parameter space, they also require specific and unwieldly selections of field content, with ten or more joint scalars, and up to nineteen dark $\rep{1}{3}$ scalars\footnote{See Fig.~10 of Ref.~\cite{Ritter:2022opo}.}.

To summarise the conclusions of our previous work regarding the FCFP approach:
\begin{itemize}
    \item For a model to have a perturbative FCFP, there must be a precise cancellation between combinations of different $\beta$-function coefficients. In addition, to obtain a decently large value for $\epsilon_v$, the couplings at the fixed point must be small. This means that only a very small proportion of models ($\lesssim 1\%$) even have such a fixed point.
    \item To then have $M >$~TeV in most of the UV-coupling parameter space, we need the couplings to run strongly in the full theory while keeping the one-loop $\beta$-function coefficients negative to ensure the existence of an infrared FCFP. This requires large multiplicities for certain fields, e.g. jointly-charged $\rep{3}{3}$ scalars.
\end{itemize}


\section{The zero-coupling infrared fixed point approach}\label{sec:zero_coupling}

In this section we consider models in which the zero-coupling fixed point of Eqn.~\ref{eqn:zcfp_solution}, $\alpha_s^\star = \alpha_d^\star = 0$, is an infrared fixed point. This is intended to address the issues with the FCFP approach identified in the previous section, with the benefits being as follows:
\begin{itemize}
    \item A large proportion of models have an infrared ZCFP, and so viable models can be found more generically within this framework.
    \item Models including fields in higher-dimensional representations of $SU(3)_\mathrm{QCD} \times SU(N_d)_\mathrm{dQCD}$ achieve infrared ZCFPs with small field multiplicities.
    \item In these models, the running of the gauge couplings is generically strong in the full theory, such that the mass scale of new physics $M$ is well above the TeV scale.
\end{itemize}

In the previous section it was noted that FCFPs with small coupling values produced the largest viability fractions, with an intuitive explanation being that the initial UV couplings $\alpha_s^\mathrm{UV}$ and $\alpha_d^\mathrm{UV}$ are `focused' towards the small fixed point, and thus brought close to each other in value at lower energy scales. A zero-coupling infrared fixed point has the same `focusing' effect on the gauge couplings, and so it is reasonable to expect that models with such an IRFP may have decent values of $\epsilon_v$. 

However, this approach comes with an inherent issue that we have already mentioned: the inclusion of Dirac fermions and complex scalars with large masses introduces a naturalness issue, as they contribute to the running of the Higgs mass parameter. These contributions become significant for new physics with masses in the 10s--100s of TeV~\cite{Clarke:2016jzm}, while in Section~\ref{sec:models} we generally consider models with $M \gtrsim 10^{15}$ GeV. For such models, we are thus led to incorporate supersymmetry within the framework.

\subsection{Obtaining a zero-coupling infrared fixed point}

The zero-coupling point $\alpha_s^\star = \alpha_d^\star = 0$ is trivially always a fixed point of the RG evolution. For it to be an infrared fixed point, the $\beta$-functions must be positive for gauge couplings in the vicinity of the fixed point: that is, for small couplings $\alpha_s, \alpha_d \ll 1$. In this regime, the one-loop terms dominate, and so the one-loop $\beta$-function coefficients must be positive:
\begin{equation}
    b_v^{(0)}, b_d^{(0)} > 0.
\end{equation}

This is a much less restrictive requirement than the delicate cancellations that we needed to obtain an infrared FCFP. Since matter contributions to the $\beta$-function coefficients are positive, all a model needs in order to have an infrared ZCFP is a sufficient amount of new field content such that the one-loop terms become positive.

In addition, infrared ZCFPs can be achieved generically with low multiplicities for the new field content, thanks to the presence of fields in higher-dimensional representations of both visible and dark QCD. These fields contribute more strongly to the $\beta$-function coefficients in Eqn.~\ref{eqn:full_beta}, due to their high dimension and larger group theory factors. So, for example, if $N_d = 3$, a model with 6 quarks, 1 dark quark, and just a single $\rep{6}{6}$ fermion exhibits an infrared ZCFP.

To see which higher-dimensional representations have this effect of obtaining infrared ZCFPs with low multiplicities, we consider models containing the 6 SM quarks, 1 dark quark, and some number of copies of an $\RvRd$ fermion or scalar. In Table~\ref{tab:zcfp_multiplicities} we give the minimum number of copies of this field that must be included for a model to exhibit an infrared ZCFP: that is, for the one-loop $\beta$-function coefficients $b_v^{(0)}$ and $b_d^{(0)}$ to be positive. Fermions of a given representation require smaller multiplicities than the corresponding scalar, since the fermions make a larger contribution to each $\beta$-function coefficient; this is due in part to Dirac fermions having more degrees of freedom than complex scalars. We also note that $N_d = 4$ models require more higher-dimensional representations than $N_d = 2, 3$, which correlates with a larger negative gluonic contribution to the $\beta$-function coefficients.

\begin{table}[]
\centering
\begin{tabular}{c|ccc|ccc|ccc}
$N_d$          & \multicolumn{3}{c}{2}                & \multicolumn{3}{c}{3}                & \multicolumn{3}{c}{4}                 \\ \hline
               & \multicolumn{9}{c}{$\mathbf{R}_d$}   \\ \cline{2-10} 
$\mathbf{R}_v$ & \textbf{2} & \textbf{3} & \textbf{4} & \textbf{3} & \textbf{6} & \textbf{8} & \textbf{4} & \textbf{6} & \textbf{10} \\ \hline
\textbf{3}     & 6          & 4          & 3          & 6          & 2          & 2          & 7          & 4          & 2           \\
\textbf{6}     & 2          & 1          & 1          & 3          & 1          & 1          & 4          & 2          & 1           \\
\textbf{8}     & 2          & 1          & 1          & 2          & 1          & 1          & 3          & 2          & 1           \\ \hline
\multicolumn{10}{c}{\textbf{(a) fermion}}\\
\multicolumn{10}{c}{}
\end{tabular}

\begin{tabular}{c|ccc|ccc|ccc}
$N_d$          & \multicolumn{3}{c}{2}                & \multicolumn{3}{c}{3}                & \multicolumn{3}{c}{4}                 \\ \hline
               & \multicolumn{9}{c}{$\mathbf{R}_d$}   \\ \cline{2-10} 
$\mathbf{R}_v$ & \textbf{2} & \textbf{3} & \textbf{4} & \textbf{3} & \textbf{6} & \textbf{8} & \textbf{4} & \textbf{6} & \textbf{10} \\ \hline
\textbf{3}     & 22         & 15         & 11         & 21         & 8          & 6          & 28         & 14         & 5           \\
\textbf{6}     & 7          & 3          & 3          & 11         & 3          & 2          & 14         & 7          & 3           \\
\textbf{8}     & 5          & 3          & 2          & 8          & 2          & 2          & 11         & 6          & 2           \\ \hline
\multicolumn{10}{c}{\textbf{(b) scalar}}
\end{tabular}
\caption{Each position in a table specifies a representation $\RvRd$ of $SU(3)_\mathrm{QCD} \times SU(N_d)_\mathrm{dQCD}$; note that the possible dark representations $\mathbf{R}_d$ change depending on the value of $N_d$. The table entries give the minimum multiplicity of a Dirac fermion \textbf{(top)} or complex scalar \textbf{(bottom)} in that representation that must be included in a model containing the 6 SM quarks and 1 dark quark, such that the model exhibits a zero-coupling infrared fixed point.}\label{tab:zcfp_multiplicities}
\end{table}

\subsection{Asymptotic freedom}

One may be perturbed by the fact that positive one-loop $\beta$-function coefficients seem to imply the loss of asymptotic freedom. Indeed, since the gauge couplings grow with increasing energy scale, they diverge at some Landau pole, which we must ensure does not fall below the Planck scale.

We first note that in this framework, $b_v^{(0)}$ and $b_d^{(0)}$ are only positive above the mass scale of new physics. At energies below this scale, when the new heavy fields have decoupled, the running of $\alpha_s$ is SM-like, and the running of $\alpha_d$ is similar so long as $n_{dq}$ is not too large; so, the theory is asymptotically free at low energies.

The values of the gauge couplings at high energies, and thus the location of the Landau pole, depend upon the mass scale $M$ at which the RG evolution changes. In our calculations, we begin by specifying a pair of (perturbative) gauge couplings at the UV scale, $(\alpha_s^\mathrm{UV}, \alpha_d^\mathrm{UV})$, and then solve for $M$ by matching $\alpha_s$ to its measured value at $M_Z$. This ensures that we only ever consider mass scales $M$ for which each coupling is perturbative from its associated confinement scale all the way up to the Planck scale.

Indeed, the philosophy behind our idea of a model that `naturally' explains the coincidence problem is that we are agnostic to physics above the UV scale. We thus treat $\alpha_s^\mathrm{UV}$ and $\alpha_d^\mathrm{UV}$ as input parameters, and when defining the viability fraction $\epsilon_v$ we consider them to take perturbative values that are randomly, uniformly and independently chosen. This `flat prior' may well be invalid,  as the `initial' gauge coupling values in the UV are in fact generated by the unknown high-scale physics. This physics above the Planck scale may also change the RG evolution, avoiding a Landau pole in the theory by either regaining asymptotic freedom, or by the couplings approaching an UV fixed point as in the `asymptotic safety' idea~\cite{Weinberg:1976xy, Weinberg:1980gg, Bonanno:2020bil}. But, given our ignorance of specific Planck-scale scenarios for generating the coupling constant values, a flat prior is justified.

\subsection{The mass scale of new physics}

Models with an infrared ZCFP generally require a high mass scale for the new physics. In these models, the positive $\beta$-function coefficients in the full theory lead to strong running of the gauge couplings from their initial values in the UV, unless there is a precise cancellation such that the $\beta$-function coefficients happen to be small. As a result of this `fast' RG evolution, to generate the correct value for $\alpha_s$ at $M_Z$, the new physics must generically decouple at a high scale.

We illustrate this analytically at one-loop level. Ignoring threshold corrections, for a given value of the strong coupling in the UV, $\alpha_s^\mathrm{UV}$, the mass scale $M$ for the new physics is found by running $\alpha_s$ up from $\alpha_s(M_Z)$ using the SM RGE, and down from $\alpha_s^\mathrm{UV}$ using the RGE of the full theory until they meet. At one-loop level, this is given by
\begin{multline}
    \alpha_s^{-1}(M_Z) - b_5\log\left(\frac{m_t}{M_Z}\right) - b_6\log\left(\frac{M}{m_t}\right) \\
    =  (\alpha_s^\mathrm{UV})^{-1} - b\log\left(\frac{M}{M_\mathrm{Pl}}\right),
\end{multline}
where $b_n = (2n/3 - 11)/2\pi$ is the one-loop $\beta$-function coefficient for QCD with $n$ quark flavours, and $b \equiv b_v^{(0)}/2\pi$ is the one-loop coefficient for the full theory. Taking the maximum value of $\alpha_s^\mathrm{UV} = 0.3$ gives the minimum value for the mass scale as
\begin{equation}
    M \sim M_\mathrm{Pl}\exp\left(-\frac{56}{b+1.1}\right).
\end{equation}

While the resultant mass scale can be low, this requires $b$ to be small. As will be discussed shortly, $M$ cannot be more than 100s of TeV without introducing a severe naturalness issue. To obtain such a mass scale,
\begin{equation}
    M \sim 1\text{---}100\text{~TeV} \implies b \sim 0.4\text{---}0.6.
\end{equation}

For this to occur, the matter contribution to $b$ must cancel closely with the gluonic contribution. We encountered a similar restriction in the FCFP approach when searching for models with a sufficiently large value of $M$, and in that case we required specific and unwieldy selections of field content. We find similar behaviour in this case, which we explore in Section~\ref{sec:lowM_models}.

We identified earlier that we would be interested in models containing fields in higher-dimensional representations, as these models generically obtain an infrared ZCFP with low multiplicities for the new fields. However, it is the inclusion of such representations that strengthens the running to the point that the mass scale is very high. For example, for $N_d = 3$, models including a $\rep{6}{6}$ fermion have $b \gtrsim 2$, leading to $M \gtrsim 10^{11}$~GeV. 

In Table~\ref{tab:min_M} we show the minimum mass scales of new physics for models including a single fermion in the representation $\RvRd$. We only show the representations for which a model with the 6 SM quarks, 1 dark quark, and 1 $\RvRd$ fermion has an infrared ZCFP: that is, the representations in Table~\ref{tab:zcfp_multiplicities} with minimum multiplicity 1. We reiterate that these are only the minimum values for $M$, calculated at one-loop; using the two-loop RGEs, working with any UV coupling below $\alpha_s^\mathrm{UV} = 0.3$, and including any additional field content in these models only increases $M$.

\begin{table}[]
\begin{tabular}{c|cc|cc|c}
$N_d$          & \multicolumn{2}{c}{2}                & \multicolumn{2}{c}{3}                & 4                 \\ \hline
               & \multicolumn{5}{c}{$\mathbf{R}_d$}   \\ \cline{2-6} 
$\mathbf{R}_v$ & \textbf{3} & \textbf{4} & \textbf{6} & \textbf{8} & \textbf{10} \\ \hline
\textbf{6}     & $10^3$          & $10^7$          & $10^{11}$          & $10^{13}$          & $10^{14}$           \\
\textbf{8}     & $10^6$          & $10^9$          & $10^{12}$          & $10^{14}$          & $10^{15}$           \\ \hline
\end{tabular}
\caption{Each position in the table specifies a representation $\RvRd$ of $SU(3)_\mathrm{QCD} \times SU(N_d)_\mathrm{dQCD}$ as before, except we have only included the representations for which a model only requires a single fermion to generate an infrared ZCFP (i.e. the `1' entries in Table~\ref{tab:zcfp_multiplicities}). The table entries give the order of magnitude for the minimum mass $M$ in GeV of a fermion in that representation, assuming $\alpha_s^\mathrm{UV} = 0.3$ and calculated at one-loop level.}\label{tab:min_M}
\end{table}

\subsection{Naturalness and a heavy mass scale}

We now come to the naturalness issue that arises due to the presence of heavy new physics. As we have seen, the ZCFP approach to this framework generally necessitates such heavy fields.

Dirac fermions and complex scalars contribute to the running of the Higgs mass parameter. If these fields are sufficiently heavy, then these contributions are sizeable, making its low-energy value $\mu^2(M_Z)$ very sensitive to physics at a high scale. This is true even for coloured fermions, which only contribute to the Higgs mass parameter at three-loop level.

To assess the scale of this issue, we adopt the approach of Ref.~\cite{Clarke:2016jzm}, where a model is unnatural if, given a flat prior for the Higgs mass parameter in the UV, it is very unlikely for it to obtain its measured value at $M_Z$. This is similar in spirit to our perspective on naturalness, but is implemented quantitatively using a Bayesian sensitivity measure that can be roughly interpreted as a level of fine-tuning.

In Ref.~\cite{Clarke:2016jzm}, limits are placed on the masses of new gauge multiplets added to the SM. These limits are calculated for threshold values of the sensitivity measure; we look at the results corresponding to a 10\% fine-tuning, where for colour-triplet, electroweak-singlet Dirac fermions, the limits are in the 10---100~TeV mass range, and for colour-triplet, electroweak-singlet complex scalars, the limits are in the 1---10~TeV mass range.

We note that these limits on $M$ may be optimistic for the models we consider. Firstly, they only apply to colour triplets; any field in a higher-dimensional representation of $SU(3)_\mathrm{QCD}$ will be subject to a stronger constraint. Secondly, these limits are calculated for models that only have a single additional gauge multiplet. In models where multiple new fermions and scalars are introduced, $M$ must be smaller again to avoid excessive fine-tuning. For these reasons it is possible that even for models with $M \lesssim 1$~TeV the naturalness issue could remain problematic. This suggests that it could even be impossible to find models in this framework that avoid a naturalness problem, an observation which also applies to the FCFP approach.

\subsection{Supersymmetry}

To consider models in which the mass scale $M$ is much greater than 100s of TeV, we introduce supersymmetry, which prevents the new heavy fields from contributing to the running of the Higgs mass parameter. We always take the scale of supersymmetry breaking to be $M_\mathrm{SUSY} = 5$~TeV. Below $M_\mathrm{SUSY}$ the running is as before, with the running of $\alpha_d$ due to $n_{dq}$ light dark quarks. Between $M_\mathrm{SUSY}$ and $M$ the gauge couplings run separately; the RG evolution of $\alpha_s$ is MSSM-like, and the RG evolution of $\alpha_d$ is due to $n_{dq}$ light dark quarks and $n_{dq}$ light dark squarks.

The heavy field content in this case consists of chiral supermultiplets. In the non-SUSY case, we introduce Dirac fermions and complex scalars to ensure anomaly cancellation and write down renormalisable mass terms. For the same reasons, in the SUSY case when we say we introduce a `field' in the representation $\RvRd$ of visible and dark QCD, we are adding two chiral supermultiplets $\Phi \sim \RvRd_y$ and $\bar{\Phi} \sim (\mathbf{\bar{R}}_v, \mathbf{\bar{R}}_d)_{-y}$ with superpotential mass term $M\bar{\Phi}\Phi$.

Introducing SUSY slightly alters the conclusions we made previously in this section about the existence of a zero-coupling IRFP and the size of the mass scale $M$ in these models.

In Table~\ref{tab:susy_zcfp_multiplicities}, we replicate Table~\ref{tab:zcfp_multiplicities} in the supersymmetric case. We find that with SUSY, even fewer copies of each new field are required for a model to exhibit a zero-coupling infrared fixed point. This is expected, as each new field we introduce is a pair of chiral supermultiplets containing the same field content as a Dirac fermion and two complex scalars.

\begin{table}[]
\begin{tabular}{c|ccc|ccc|ccc}
$N_d$          & \multicolumn{3}{c}{2}                & \multicolumn{3}{c}{3}                & \multicolumn{3}{c}{4}                 \\ \hline
               & \multicolumn{9}{c}{$\mathbf{R}_d$}   \\ \cline{2-10} 
$\mathbf{R}_v$ & \textbf{2} & \textbf{3} & \textbf{4} & \textbf{3} & \textbf{6} & \textbf{8} & \textbf{4} & \textbf{6} & \textbf{10} \\ \hline
\textbf{3}     & 2          & 2          & 1          & 3          & 1          & 1          & 4          & 2          & 1           \\
\textbf{6}     & 1          & 1          & 1          & 2          & 1          & 1          & 2          & 1          & 1           \\
\textbf{8}     & 1          & 1          & 1          & 1          & 1          & 1          & 2          & 1          & 1           \\ \hline
\end{tabular}
\caption{The same as Table~\ref{tab:zcfp_multiplicities}, but incorporating supersymmetry. Each table entry now gives the minimum multiplicity of pairs of chiral supermultiplets in a given representation that must be included in a model with the 6 SM (s)quarks and 1 dark (s)quark, such that the model exhibits an infrared ZCFP.}\label{tab:susy_zcfp_multiplicities}
\end{table}

We also find that the required mass scale $M$ is larger when SUSY is included. For a given $\alpha_s^\mathrm{UV}$, at one-loop level and ignoring threshold corrections, $M$ is found by
\begin{multline}
    \alpha_s^{-1}(M_Z) - b_5\log\left(\frac{m_t}{M_Z}\right)\\ - b_6\log\left(\frac{M_\mathrm{SUSY}}{m_t}\right) - b_\mathrm{SUSY}\log\left(\frac{M}{M_\mathrm{SUSY}}\right)\\
    =  (\alpha_s^\mathrm{UV})^{-1} - b\log\left(\frac{M}{M_\mathrm{Pl}}\right),
\end{multline}
where the the one-loop $\beta$-function coefficient $b_\mathrm{SUSY} = -3/2\pi$. Since the MSSM running is weaker than the SM running, to match the measured strong coupling $\alpha_s(M_Z)$, the heavy fields must decouple at a higher scale. With $M_\mathrm{SUSY} = 5$~TeV and again taking $\alpha_s^\mathrm{UV} = 0.3$, the minimum value for the mass scale in the SUSY case is
\begin{equation}
    M \sim M_\mathrm{Pl}\exp\left(-\frac{40}{b+0.5}\right).
\end{equation}

In Table~\ref{tab:susy_min_M} we replicate Table~\ref{tab:min_M} for the SUSY case, showing the minimum mass scales of new physics for models including a single field in the representation $\RvRd$. As before, we only show the values of $M$ for representations where a model with the 6 SM (s)quarks, 1 dark (s)quark, and 1 $\RvRd$ field has an infrared ZCFP. Since this occurs for lower-dimensional representations in the SUSY case, more values of $M$ are shown than in Table~\ref{tab:min_M}. There are also two representations marked with crosses; these models do happen to have a small value of $b$, and so in fact have $M < M_Z$. This is not possible, and indicates that one cannot solve for $M$ in such a model for $\alpha_s^\mathrm{UV} = 0.3$.

As a comparison to the point singled out in the non-SUSY case, for $N_d = 3$, models including a $\rep{6}{6}$ field have $b \gtrsim 4$, leading to $M \gtrsim 10^{15}$~GeV. 

Finally, we note that although these heavy superfields do not introduce a naturalness issue, we still include their mass scales as they are relevant for some of the discussion in Section~\ref{sec:adm}.

\begin{table}[]
\begin{tabular}{c|ccc|ccc|ccc}
$N_d$          & \multicolumn{3}{c}{2}                & \multicolumn{3}{c}{3}                & \multicolumn{3}{c}{4}                 \\ \hline
               & \multicolumn{9}{c}{$\mathbf{R}_d$}   \\ \cline{2-10} 
$\mathbf{R}_v$ & \textbf{2} & \textbf{3} & \textbf{4} & \textbf{3} & \textbf{6} & \textbf{8} & \textbf{4} & \textbf{6} & \textbf{10} \\ \hline
\textbf{3}     & -          & -          & $\times$          & -          & $\times$          & $10^5$          & -          & -          & $10^8$           \\
\textbf{6}     & $10^8$          & $10^{12}$          & $10^{13}$          & -          & $10^{15}$          & $10^{16}$          & -          & $10^{15}$          & $10^{17}$           \\
\textbf{8}     & $10^{10}$          & $10^{13}$          & $10^{14}$          & $10^{13}$          & $10^{16}$          & $10^{17}$          & -          & $10^{16}$          & $10^{17}$           \\ \hline
\end{tabular}
\caption{The same as Table~\ref{tab:min_M}, but incorporating supersymmetry. Dashes indicate that more than one copy of the given field is required to generate an infrared ZCFP. Crosses denote models for which the minimum mass is below $M_Z$.}\label{tab:susy_min_M}
\end{table}


\section{Finding viable models}\label{sec:models}

In this section we identify models within the zero-coupling fixed point approach that can viably explain the cosmological coincidence problem; that is, models with a sufficiently large value of $\epsilon_v$.

We first consider models without supersymmetry, where we must ensure that the mass scale of new physics $M$ is in the 1---100~TeV range to avoid introducing a naturalness issue. We find some successful models, but also illustrate the delicate cancellations required in obtaining this low mass scale.

We then look at models incorporating supersymmetry, where a high mass scale is allowed. We find that the viability of a given model depends strongly on the number of light dark quarks, and analyse sets of models for $N_d = 2$, 3, and 4. We find the most promising results for $N_d = 4$ where most models ($\sim$65\%) have a decently large viability fraction ($\epsilon_v > 0.3$). For $N_d = 3$, we find that a decent proportion of models ($\sim$~40\%) can obtain a decently large viability fraction ($\epsilon_v > 0.25$), and for $N_d = 2$ we find that less than 1\% of models achieve $\epsilon_v > 0.25$.

\subsection{Viable models with a low mass scale}\label{sec:lowM_models}

To find models with a low mass scale of new physics $M$, we first must specify the upper limit on $M$ such that we avoid a naturalness issue from the contribution of these new fields to the running of the Higgs mass parameter. As mentioned in the previous section, we are guided by the results of Ref.~\cite{Clarke:2016jzm} which provides limits on the mass of new Dirac fermion and complex scalar gauge multiplets. Conservatively, to introduce less than 10\% fine-tuning requires:
\begin{itemize}
    \item $M \lesssim 100$~TeV for colour-triplet, electroweak-singlet Dirac fermions.
    \item $M \lesssim 10$~TeV for colour-triplet, electroweak-singlet complex scalars.
\end{itemize}

We use these limits as rough criteria for determining whether a model avoids a naturalness problem. For models containing only new fermions we look at the viable proportion of $(\alpha_s^\mathrm{UV}, \alpha_d^\mathrm{UV})$ parameter space for which $M < 100$~TeV: we call this parameter $\epsilon_v^{100}$. For models also containing new scalars we look at the viable proportion of UV coupling parameter space for which $M < 10$~TeV: we call this parameter $\epsilon_v^{10}$.

So, what conditions do we need to place on a model to have $M \lesssim 10\text{---}100$~TeV in a sizeable proportion of the UV coupling parameter space?

In the previous section, to one-loop level we found that a one-loop $\beta$-function coefficient $b_v^{(0)}/2\pi \sim 0.4\text{---}0.6$ results in a minimum value of $M$ between 1 and 100 TeV. This changes when we include the two-loop RG evolution. At two-loop level the running is stronger, and so to obtain the same value of $M$ requires a smaller value for $b_v^{(0)}/2\pi$. The value of $M$ now also depends on $\alpha_d^\mathrm{UV}$, and so for a given model the minimum mass scale $M_\mathrm{min}$ occurs for $(\alpha_s^\mathrm{UV}, \alpha_d^\mathrm{UV}) = (0.3, 0)$. We now find that
\begin{align}
    &M_\mathrm{min} \lesssim 10\text{~TeV} \implies \frac{b_v^{(0)}}{2\pi} \lesssim 0.1, \\
    &M_\mathrm{min} \lesssim 100\text{~TeV} \implies \frac{b_v^{(0)}}{2\pi} \lesssim 0.2.
\end{align}

\subsubsection{Models including scalars}

Let us first consider models including new complex scalars, where we are interested in the region of viable UV coupling parameter space with $M < 10$~TeV. Clearly $b_v^{(0)}/2\pi \lesssim 0.1$ is a necessary requirement for such a region to exist; however, it does not ensure that this region will be large. Indeed, we find that a decent value for $\epsilon_v^{10}$ can only be obtained for the smallest possible value of $b_v^{(0)}$.

We show this by considering three example models with different values of $b_v^{(0)}/2\pi < 0.1$. We first note that the smallest contribution to the one-loop $\beta$-function coefficient comes from the $\rep{3}{1}$ scalar, and is equal to $1/12\pi \sim 0.03$; all other fields contribute some multiple of this amount. So, there are are only three possible values of $b_v^{(0)}/2\pi < 0.1$. In Table~\ref{tab:lowM_lowb_models} we give the field content of the three example models, along with their value of $b_v^{(0)}/2\pi$ and the minimum mass scale of new physics $M_\mathrm{min}$.

\begin{table}[]
\begin{tabular}{ccc|cccc}
    \textbf{fermions} & $n_{dq}$ & \textbf{scalars} & $b_v^{(0)}/2\pi$ & $M_\mathrm{min}$ & $\epsilon_v$ & $\epsilon_v^{10}$ \\ \hline
    $\begin{pmatrix}
       & 6\\
     6 & 3
    \end{pmatrix}$
    & 4 &
    $\begin{pmatrix}
       & 1\\
     1 & 2
    \end{pmatrix}$
    & $1/12\pi$ & 1.1 & 0.35 & 0.27\\
    $\begin{pmatrix}
       & 6\\
     6 & 3
    \end{pmatrix}$
    & 5 &
    $\begin{pmatrix}
       & 1\\
     2 & 2
    \end{pmatrix}$
    & $1/6\pi$ & 2.2 & 0.34 & 0.15\\
    $\begin{pmatrix}
       & 6\\
     6 & 3
    \end{pmatrix}$
    & 6 &
    $\begin{pmatrix}
       & 1\\
     3 & 2
    \end{pmatrix}$
    & $1/4\pi$ & 4.2 & 0.33 & 0\\
\end{tabular}
\caption{Three models with low values of $M$. For each model we give the one-loop $\beta$-function coefficient $b_v^{(0)}/2\pi$, the minimum scale of new physics $M_\mathrm{min}$ in TeV, the viability fraction $\epsilon_v$, and the viable proportion of UV coupling parameter space for which $M < 10$~TeV, $\epsilon_v^{10}$. The field content of each model charged under QCD and dark QCD is specified in the three left-most columns. In the fermion (scalar) matrices, each entry gives the multiplicity of a Dirac fermion (complex scalar) in a given representation $\RvRd$ of $SU(3)_\mathrm{QCD}\times SU(3)_\mathrm{dQCD}$. The rows correspond to $\mathbf{R}_v = \mathbf{1}, \mathbf{3}$ from top to bottom, and the columns correspond to $\mathbf{R}_d = \mathbf{1}, \mathbf{3}$ from left to right. The fact that no further rows or columns are given indicates that no fields in higher-dimensional representations are present. While the top-right entry of the fermion matrix gives the total number of $\rep{1}{3}$ fermions, some are at the heavy mass scale $M$, and some have light masses below $\Lambda_\mathrm{dQCD}$; so, we also specify the number of light dark quarks, $n_{dq}$. The contours for $M$ and $\Lambda_\mathrm{dQCD}$ in $(\alpha_s^\mathrm{UV}, \alpha_d^\mathrm{UV})$ parameter space for each model are shown in Fig.~\ref{fig:lowM_lowb_models}.}
\label{tab:lowM_lowb_models}
\end{table}

\begin{figure*}
    \centering
    \includegraphics[width=0.3\textwidth]{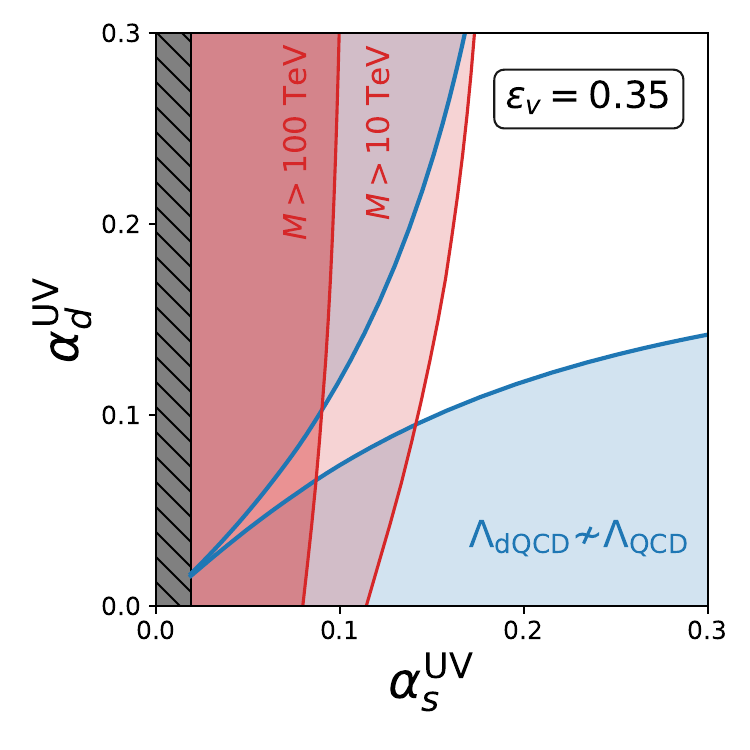}\includegraphics[width=0.3\textwidth]{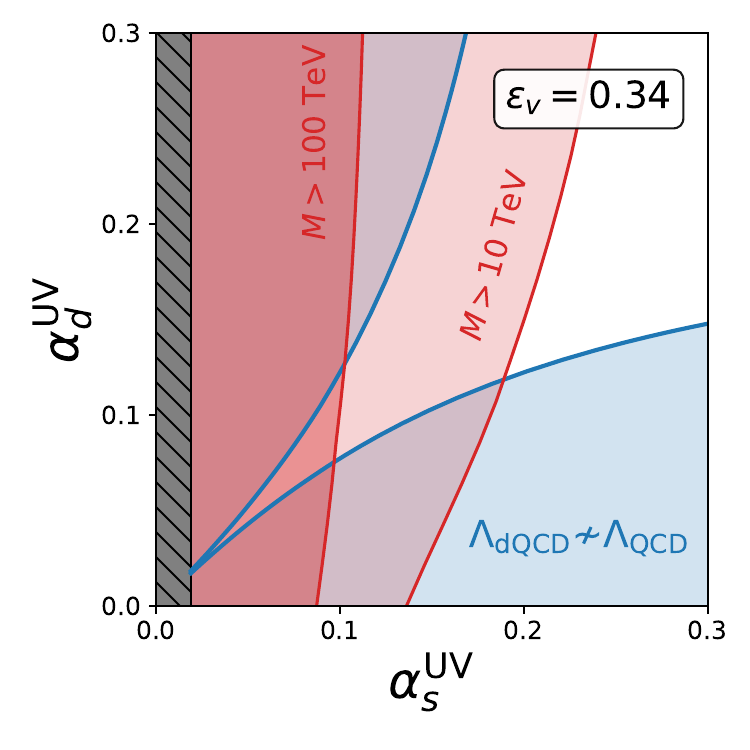}\includegraphics[width=0.3\textwidth]{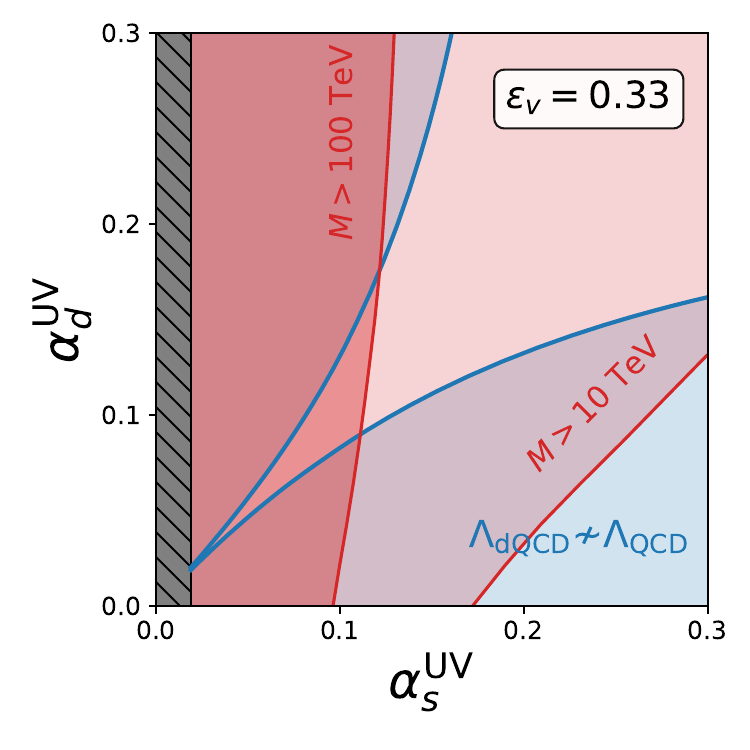}
    \caption{The UV coupling parameter space for three models with low values of $M$. The field content for each model is given in Table~\ref{tab:lowM_lowb_models}. From left to right, the red contours show the UV couplings for which $M = 100$~TeV and $M = 10$~TeV. The upper (lower) blue contour shows the UV couplings for which $\Lambda_\mathrm{dQCD} = 5$~(0.2)~GeV. In the grey hatched region, there is no value of $M$ for which the measured value of $\alpha_s(M_Z)$ can be reproduced.}
    \label{fig:lowM_lowb_models}
\end{figure*}

In Table~\ref{tab:lowM_lowb_models} we also give the value of $\epsilon_v$  and $\epsilon_v^{10}$ for each model. Only the model with the smallest $b_v^{(0)}/2\pi$ obtains a decent value for $\epsilon_v^{10}$. In Fig.~\ref{fig:lowM_lowb_models} we show the relevant $M$ and $\Lambda_\mathrm{dQCD}$ contours in $(\alpha_s^\mathrm{UV}, \alpha_d^\mathrm{UV})$ parameter space for these models; in these plots, $\epsilon_v$ corresponds to the region between the blue curves and $\epsilon_v^{10}$ corresponds to the unshaded region.

To obtain such a small value for $b_v^{(0)}$ requires a specific cancellation of the contributions from the gauge and matter content. We would like to achieve this cancellation with low multiplicities for the new fields. To quantify the notion of how rarely these cancellations occur, we consider a simple set of models with at most three of each new field. Since the value of $b_v^{(0)}$ only depends on fields charged under QCD, in these models we only specify the multiplicity of such fields.

Including at most triplet fields, there are 256 total models, of which 5 ($\sim$2\%) have $b_v^{(0)}/2\pi = 1/12\pi$. Including up to sextets, there are now $\sim$16 million models, of which only 242 ($\sim$0.001\%) have $b_v^{(0)}/2\pi = 1/12\pi$.

In these models we have not specified the dark field content -- that is, fields charged only under $SU(N_d)_\mathrm{dQCD}$. Changing the dark field content changes $\epsilon_v$ and $\epsilon_v^{10}$; so, to find viable models with a sufficiently low value of $M$ we add dark fields until $\epsilon_v^{10}$ is decently large. For a given selection of field content charged under QCD, there may be multiple selections of dark field content for which this occurs. In Table~\ref{tab:lowM_maxtriplet_models}, we give one such example for each of the 5 models with only new triplet fields, where we have added in a number of light and heavy dark quarks.

When adding dark field content to the 242 models including sextets, we find that most of them are not able to obtain a decent value for $\epsilon_v^{10}$. This occurs for two reasons. Firstly, despite these models having the smallest possible value for the one-loop $\beta$-function coefficient $b_v^{(0)}$, if the two-loop coefficients are too large, then $M < 10$~TeV in only a small region of the UV coupling parameter space. Secondly, if most of the field content is in a higher-dimensional representation of dark QCD than visible QCD, then $b_d^{(0)}$ is larger than $b_v^{(0)}$ even before the addition of any dark field content.  The result of this is that $\epsilon_v$ is very small, as the stronger running of $\alpha_d$ than $\alpha_s$ in the UV means that similar confinement scales will only be obtained when $\alpha_d^\mathrm{UV} \gg \alpha_s^\mathrm{UV}$. This issue only worsens with the inclusion of more dark field content, making $\epsilon_v$ even smaller.

Due to these additional difficulties, we find that only 16 of the 242 models with sextets obtain a decent value for $\epsilon_v^{10}$. We show the field content of these models in Appendix~\ref{sec:sextet_model_appendix}. So, while we find that there do exist models with a low enough value of $M$ to avoid a naturalness issue, we are in a similar situation to the FCFP approach: these models make up only a very small proportion of the total models, and require several stringent conditions to be satisfied. However, unlike the FCFP approach, these models do not require very large multiplicities for any of the new fields; this makes it more feasible to use these models for further model-building within a full theory of asymmetric dark matter.

\begin{table}[]
\begin{tabular}{ccc|cc}
    \textbf{fermions} & $n_{dq}$ & \textbf{scalars} & $\epsilon_v$ & $\epsilon_v^{10}$ \\ \hline
    $\begin{pmatrix}
       & 7\\
     6 & 3
    \end{pmatrix}$
    & 1 &
    $\begin{pmatrix}
       & 0\\
     1 & 2
    \end{pmatrix}$ & 0.36 & 0.30\\
    $\begin{pmatrix}
       & 7\\
     7 & 3
    \end{pmatrix}$
    & 4 &
    $\begin{pmatrix}
       & 0\\
     0 & 1
    \end{pmatrix}$ & 0.36 & 0.30\\
    $\begin{pmatrix}
       & 8\\
     7 & 3
    \end{pmatrix}$
    & 3 &
    $\begin{pmatrix}
       & 0\\
     3 & 0
    \end{pmatrix}$ & 0.37 & 0.33\\
    $\begin{pmatrix}
       & 9\\
     8 & 2
    \end{pmatrix}$
    & 2 &
    $\begin{pmatrix}
       & 0\\
     2 & 3
    \end{pmatrix}$ & 0.36 & 0.29\\
    $\begin{pmatrix}
       & 9\\
     9 & 2
    \end{pmatrix}$
    & 5 &
    $\begin{pmatrix}
       & 0\\
     1 & 2
    \end{pmatrix}$ & 0.35 & 0.28\\
\end{tabular}
\caption{Considering only the field content with $\mathbf{R}_v = \mathbf{3}$, these are the only five models with $b_v^{(0)}/2\pi = 1/12\pi$ and at most three of each new field, where we only include fields in at most triplet representations of visible and dark QCD. For each model we have added dark quarks -- that is, both light and heavy $\rep{1}{3}$ fermions -- to achieve a decent value for $\epsilon_v^{10}$, the viable proportion of UV coupling parameter space for which $M < 10$~TeV. For each model we give the viability fraction $\epsilon_v$ as well as $\epsilon_v^{10}$. The field content of each model is specified in the same way as in Table~\ref{tab:lowM_lowb_models}.}
\label{tab:lowM_maxtriplet_models}
\end{table}

\subsubsection{Models including only fermions}

For models only including new heavy fermions, we are interested in the region of viable UV coupling parameter space with $M < 100$~TeV. This weaker limit on $M$ only really applies for fields that are colour-triplets; higher-dimensional representations of QCD will require at least $M < 10$~TeV to avoid the naturalness issue.

The weaker limit on $M$ means that these models can have a larger one-loop $\beta$-function coefficient $b_v^{(0)}/2\pi \lesssim 0.2$; this less stringent restriction should lead to a higher proportion of models with a decent value of $\epsilon_v^{100}$. However, Dirac fermions of a given representation make larger contributions to the $\beta$-function coefficients than the corresponding complex scalar, with the smallest contribution of $1/3\pi\sim0.1$ coming from the $\rep{3}{1}$ fermion. This partially offsets the advantage we gain from allowing larger values for the $\beta$-function coefficients.

To see how rare these viable models are, we again consider a set of models with at most three of each new field, only specifying the field content charged under QCD. Allowing at most triplets under QCD and sextets under dark QCD, there are 64 models, of which 6 ($\sim$10\%) have $b_v^{(0)}/2\pi \lesssim 0.2$.

As before, we add dark field content to these models to see if they can obtain a decent value of $\epsilon_v^{100}$. Only three of the models can even obtain a non-zero value for $\epsilon_v^{100}$; we show the field content of these models along with $\epsilon_v$ and $\epsilon_v^{100}$ in Table~\ref{tab:lowM_nofermion_models}. The other three models face one of the issues identified earlier: they contain too many fields in higher-dimensional representations of dark QCD, and so $b_d^{(0)}$ is larger than $b_v^{(0)}$ before any additional dark field content is introduced.

\begin{table}[]
\begin{tabular}{cc|cc}
    \textbf{fermions} & $n_{dq}$ & $\epsilon_v$ & $\epsilon_v^{100}$ \\ \hline
    $\begin{pmatrix}
       & 8 & 0\\
     8 & 3 & 0
    \end{pmatrix}$
    & 4 & 0.36 & 0.35\\
    $\begin{pmatrix}
       & 9 & 0\\
     9 & 3 & 0
    \end{pmatrix}$
    & 4 & 0.34 & 0.26\\
    $\begin{pmatrix}
       & 1 & 0\\
     9 & 1 & 1
    \end{pmatrix}$
    & 1 & 0.36 & 0.18\\
\end{tabular}
\caption{The three models with at most three of each new field (where we include fermions in at most triplet representations of visible QCD and at most sextet representations of dark QCD) such that $\epsilon_v^{100}$ is non-zero. For each model we give the viability fraction $\epsilon_v$ as well as $\epsilon_v^{100}$. The field content of each model is specified in the same way as in Table~\ref{tab:lowM_lowb_models}, but here the columns correspond to $\mathbf{R}_d = \mathbf{1}, \mathbf{3}, \mathbf{6}$ from left to right.}
\label{tab:lowM_nofermion_models}
\end{table}

So, we do find that models without a naturalness issue are less rare when only including new fermions; however, the weaker limit on $M$ in this case may well be overly optimistic, and the total number of viable models is very low.

Overall, obtaining a sufficiently low value for $M$ is achievable, and does not require large multiplicities for the new fields. However, we note that these models `only just' obtain an infrared ZCFP, having small positive values for $b_v^{(0)}$ and $b_d^{(0)}$. We thus lose part of the appeal of the ZCFP approach, where generic selections of new field content obtain infrared ZCFPs without any specific cancellations. In the next subsection, we seek to take advantage of this feature by considering supersymmetric models with high mass scales.

\subsection{Supersymmetric models}\label{sec:SUSY_models}

In models incorporating SUSY, we face no naturalness issue when adding in new fields. The mass scale of new physics $M$ can now be as high as the Planck scale, and there are no restrictions on the size of the $\beta$-function coefficients as there were in the previous section. This allows us to generically obtain infrared ZCFPs with minimal new field content; in particular we can now include fields of higher-dimensional representations which immediately ensure the existence of an infrared ZCFP.

We are interested in the values of $\epsilon_v$ that these models can obtain. The hope is that models in this framework will generically be viable; however we find that this is not always the case. This desire is primarily confronted by the fact that the value of $\epsilon_v$ strongly depends on the number of light dark quarks, $n_{dq}$. This is because changing $n_{dq}$ greatly alters the running of $\alpha_d$ following the decoupling of the new field content, which now occurs at a high mass scale. 

In this section we consider sets of models with a given heavy field content; for each selection of heavy field content, we choose the number of light dark quarks $n_{dq}$ such that $\epsilon_v$ is a maximum. This is done first for $N_d = 3$, and then for $N_d = 2$ and 4. A summary of the general findings is as follows:
\begin{itemize}
    \item For $N_d = 3$, the best value of $\epsilon_v$ is $\sim$0.35, which is decently large. However, there is a fairly wide spread of maximum values for $\epsilon_v$, and only a small proportion of models have $\epsilon_v > 0.3$. While this may challenge our desire to generically obtain viable models, $\sim$40\% of models have $\epsilon_v > 0.25$, which constitutes a decent proportion of models with a reasonable viability fraction.
    \item For $N_d = 4$, the situation improves. The best value of $\epsilon_v$ is $\sim$0.42, and $\sim$65\% of models have $\epsilon_v > 0.3$.
    \item For $N_d = 2$, the situation is worse. There are no models with $\epsilon_v > 0.3$, and less than 1\% even have $\epsilon_v > 0.25$.
    \item The required number of light dark quarks peaks at $n_{dq} = 6$ for $N_d = 3$, $n_{dq} = 11$ for $N_d = 4$, and $n_{dq} = 1$ for $N_d = 2$. For a given $N_d$, the maximum possible value of $\epsilon_v$ increases with $n_{dq}$.
\end{itemize}

\subsubsection{The dependence of $\epsilon_v$ on $n_{dq}$}

Before looking at the results for representative sets of models, we first discuss the main issue that makes it difficult for these models to generically obtain decent values of $\epsilon_v$; that is, the strong dependence of $\epsilon_v$ on the number of light dark quarks, $n_{dq}$.

\begin{figure*}
    \centering
    \includegraphics[width=0.8\textwidth]{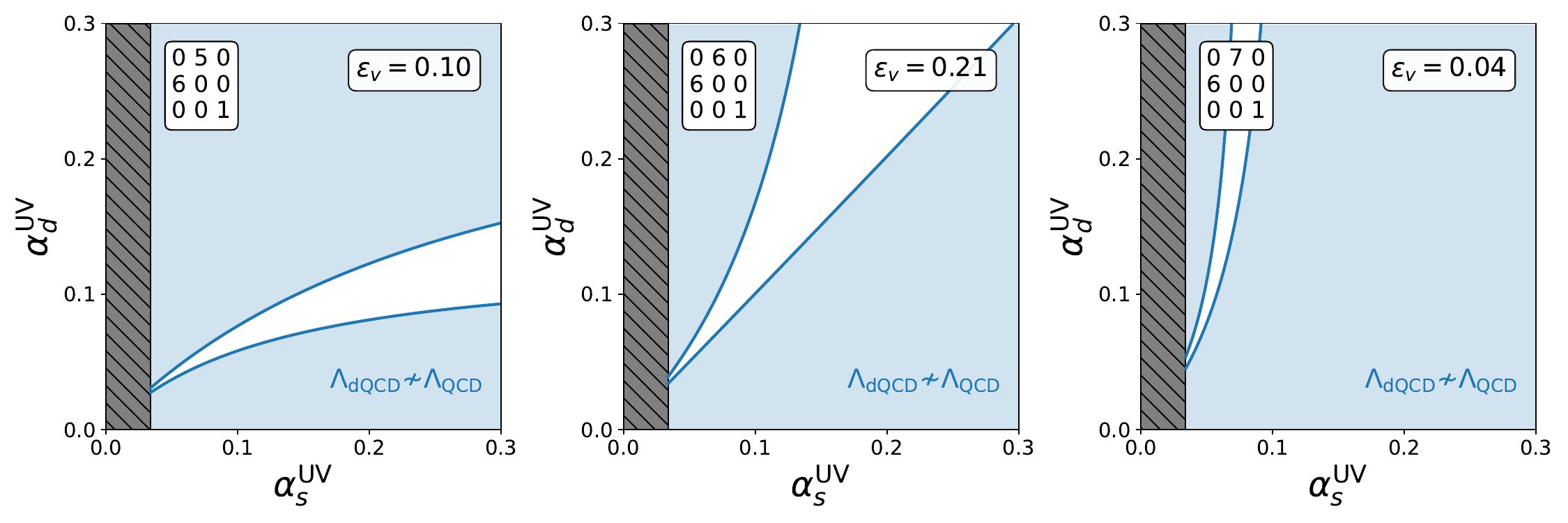}
    \caption{The UV coupling parameter space for three models with increasing values of $n_{dq}$, illustrating the strong dependence of $\epsilon_v$ on $n_{dq}$. The blue contours and grey hatched region are the same as in Fig.~\ref{fig:lowM_lowb_models}. The field content of each model is given by the matrix in the top-left corner, where each entry gives the multiplicity of a field in a given representation $\RvRd$ of $SU(3)_\mathrm{QCD}\times SU(3)_\mathrm{dQCD}$. Each field included corresponds to two chiral supermultiplets. The rows correspond to $\mathbf{R}_v = \mathbf{1}, \mathbf{3}, \mathbf{6}$ from top to bottom, and the columns correspond to $\mathbf{R}_d = \mathbf{1}, \mathbf{3}, \mathbf{6}$ from left to right. All $\rep{1}{3}$ fields are taken to have a light mass below $\Lambda_\mathrm{dQCD}$.}
    \label{fig:changing_ndq.pdf}
\end{figure*}

We note that this dependence on $n_{dq}$ occurs in all versions of this framework, including the models with FCFPs that we considered in our previous work; however, it is more sizeable for the models in this section given their large values for $M$. We illustrate this dependence in Fig.~\ref{fig:changing_ndq.pdf} where we show the region of viable UV coupling parameter space for simple models with $N_d = 3$, a single heavy $\rep{6}{6}$ field, and between 5 and 7 light dark quarks.

We first note that the viable region seems to `rotate' anti-clockwise as $n_{dq}$ increases. This is generic behaviour, and occurs because, for a given $(\alpha_s^\mathrm{UV}, \alpha_d^\mathrm{UV})$, increasing $n_{dq}$ decreases the value of $\Lambda_\mathrm{dQCD}$. The strong dependence of $\epsilon_v$ on $n_{dq}$ then comes from the fact that increasing $n_{dq}$ by 1 greatly changes the value of $\Lambda_\mathrm{dQCD}$.

To understand this, consider adding a single light dark quark to a model. This additional field only slightly affects the strong running of $\alpha_d$ in the UV, but substantially weakens the running of $\alpha_d$ following the decoupling of the new heavy field content. Also, since $M$ depends on the running of $\alpha_s$, its value stays approximately constant. So, for a given $(\alpha_s^\mathrm{UV}, \alpha_d^\mathrm{UV})$, the weaker running after decoupling leads $\alpha_d$ to reach a non-perturbative value at a lower energy scale. For strong running in the UV and a high value of $M$, a single additional light dark quark will decrease $\Lambda_\mathrm{dQCD}$ by multiple orders of magnitude.

We show this analytically at one-loop level for $N_d = 3$, assuming that $M$ and the running of $\alpha_d$ in the UV do not change when $n_{dq}$ changes. Then, for a given $\alpha_d^\mathrm{UV}$, the confinement scale for a given $n_{dq}$, $\Lambda_{\mathrm{dQCD}_n}$, is found by
\begin{multline}
    (\alpha_d^\mathrm{UV})^{-1} = \frac{4}{\pi} - b_n\log\left(\frac{M_\mathrm{SUSY}}{\Lambda_{\mathrm{dQCD}_n}}\right) \\
    - b_n^\mathrm{SUSY}\log\left(\frac{M}{M_\mathrm{SUSY}}\right) - b\log\left(\frac{M_\mathrm{Pl}}{M}\right),
\end{multline}
where $b_n = (2n_{dq}/3 - 11)/2\pi$, $b_n^\mathrm{SUSY} = (n_{dq} - 9)/2\pi$, and $b \equiv b_d^{(0)}/2\pi$ is the one-loop coefficient for the full theory. The factor by which the confinement scale changes when one light dark quark is added is then given by
\begin{equation}
    \log\left(\frac{\Lambda_{\mathrm{dQCD}_{n+1}}}{\Lambda_{\mathrm{dQCD}_n}}\right) = \frac{-A}{(33-2n_{dq})(31-2n_{dq})}
\end{equation}
with $A$ a constant given by
\begin{multline}\label{eqn:changing_ndq}
    A = 12\pi\left(b\log\left(\frac{M_\mathrm{Pl}}{M}\right) + (\alpha_d^\mathrm{UV})^{-1} - \frac{4}{\pi}\right) \\
    + 45\log\left(\frac{M}{M_\mathrm{SUSY}}\right).
\end{multline}

We see that the amount by which $\Lambda_\mathrm{dQCD}$ decreases is larger for larger $n_{dq}$. Taking typical values of $b \sim 4$ and $M \sim 10^{16}$~GeV, for $\alpha_d^\mathrm{UV} = 0.3$ we have
\begin{equation}
    \log_{10}\left(\frac{\Lambda_{\mathrm{dQCD}_{n+1}}}{\Lambda_{\mathrm{dQCD}_n}}\right) \sim -\left(1-\frac{n_{dq}}{16}\right)^{-2}.
\end{equation}
So, for $n_{dq}$ increasing from 5 to 6, we find that $\Lambda_\mathrm{dQCD}$ decreases by $\sim$2 orders of magnitude, and for $n_{dq}$ increasing from 7 to 8, we find that $\Lambda_\mathrm{dQCD}$ decreases by $\sim$3 orders of magnitude.

Finally, we note that, given the general shape of the viable regions in Fig.~\ref{fig:changing_ndq.pdf}, the models with the largest possible values for $\epsilon_v$ should be those where the top-right corner of the parameter space lies within the viable region; that is, $\Lambda_\mathrm{dQCD} \sim 1$~GeV for $(\alpha_s^\mathrm{UV}, \alpha_d^\mathrm{UV}) = (0.3, 0.3)$. This helps us understand why only a small proportion of models may have a decent value of $\epsilon_v$. Since $\Lambda_\mathrm{dQCD}$ at this point changes by multiple orders of magnitude for each additional light dark quark we add, for a given model there will not necessarily be a value of $n_{dq}$ for which the top-right corner of parameter space lies within the viable region.

\subsubsection{$N_d=3$}

With $N_d = 3$ we consider a set of selections of heavy field content where we include at most two copies of each new field, and the new fields we introduce are in the $\rep{3}{3}$, $\rep{6}{1}$, $\rep{1}{6}$, $\rep{6}{3}$, $\rep{3}{6}$, and $\rep{6}{6}$ representations of $SU(3)_\mathrm{QCD}\times SU(3)_\mathrm{dQCD}$. There are 729 such selections of field content, but we are only interested in the models that have an infrared ZCFP. Only 8 of these models lack such a fixed point, since the new field content is mostly in higher-dimensional representations of visible and dark QCD.

For each of these choices of new heavy field content, we add in $n_{dq}$ light dark quarks, choosing the value of $n_{dq}$ such that $\epsilon_v$ is a maximum.

We note that to have $\epsilon_v > 0$ requires $n_{dq} \leq 9$. This is true regardless of the field content in the model, since the maximum possible value of the dark confinement scale, $\Lambda_\mathrm{dQCD}^\mathrm{max}$, depends only on $n_{dq}$. This maximum confinement scale occurs when $\alpha_s^\mathrm{UV}$ takes its minimum value; this is the value such that $M = M_\mathrm{Pl}$, and defines the edge of the grey hatched region in each UV coupling parameter space plot. We then find $\Lambda_\mathrm{dQCD}^\mathrm{max}$ when $\alpha_d^\mathrm{UV} = 0.3$; for $n_{dq} > 9$, we have $\Lambda_\mathrm{dQCD}^\mathrm{max} < 0.2$~GeV, and so there will be no viable region in the UV coupling parameter space.

\begin{figure}
    \centering
    \includegraphics[width=0.485\textwidth]{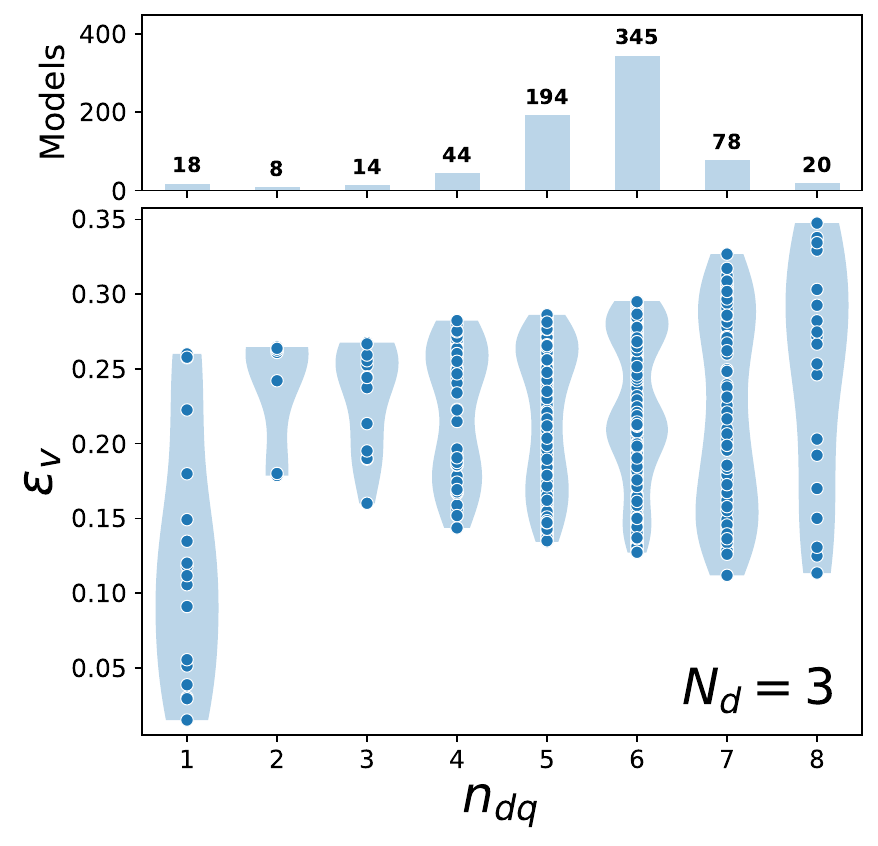}
    \caption{For a set of 721 models with $N_d = 3$, each point shows the maximum viability fraction $\epsilon_v$ and the number of light dark quarks $n_{dq}$ required to obtain it. Each model contains at most two of each new heavy field, where the fields introduced are in the $\rep{3}{3}$, $\rep{6}{1}$, $\rep{1}{6}$, $\rep{6}{3}$, $\rep{3}{6}$, and $\rep{6}{6}$ representations of $SU(3)_\mathrm{QCD}\times SU(3)_\mathrm{dQCD}$. The width of each shaded blue region shows the distribution of $\epsilon_v$ values at each $n_{dq}$. The bar chart above the main plot shows the number of models for each value of $n_{dq}$.}
    \label{fig:SUSY_Nd3_epsv_ndq.pdf}
\end{figure}

In Fig.~\ref{fig:SUSY_Nd3_epsv_ndq.pdf} we plot the maximum values of $\epsilon_v$ against $n_{dq}$ for this set of 721 models. The width of the shaded blue region gives the distribution of $\epsilon_v$ values for each $n_{dq}$. We also show the number of models for each value of $n_{dq}$; note that the widths of the shaded blue regions are not normalised by these numbers of models. We indeed see that $n_{dq} \leq 9$, and in fact no models have a maximum value of $\epsilon_v$ for $n_{dq} = 9$.

At best, we see that we have $\epsilon_v \sim 0.35$. While this is a decent value for $\epsilon_v$, at worst $\epsilon_v$ is very close to zero. Even if we discount the cluster of models with $n_{dq} = 1$ and very small viability fractions, $\epsilon_v$ can still be as low as $\sim$0.12. This challenges our desire for models in this framework to be generically viable; indeed, only 2.5\% of these models have $\epsilon_v > 0.3$, which we have previously used as a threshold for a model to be viable.

However, we do see that 39\% of models have $\epsilon_v > 0.25$. This is a reasonable viability fraction, and so we can state our result as follows: for 2 out of every 5 models in this set of models, when choosing random values for the initial gauge couplings in the UV, there is at least a 1 in 4 chance of having similar confinement scales for visible and dark QCD. We consider this a marked improvement over the coincidence problem as it arises for most canonical dark matter candidates.

\begin{figure*}
    \centering
    \includegraphics[width=\textwidth]{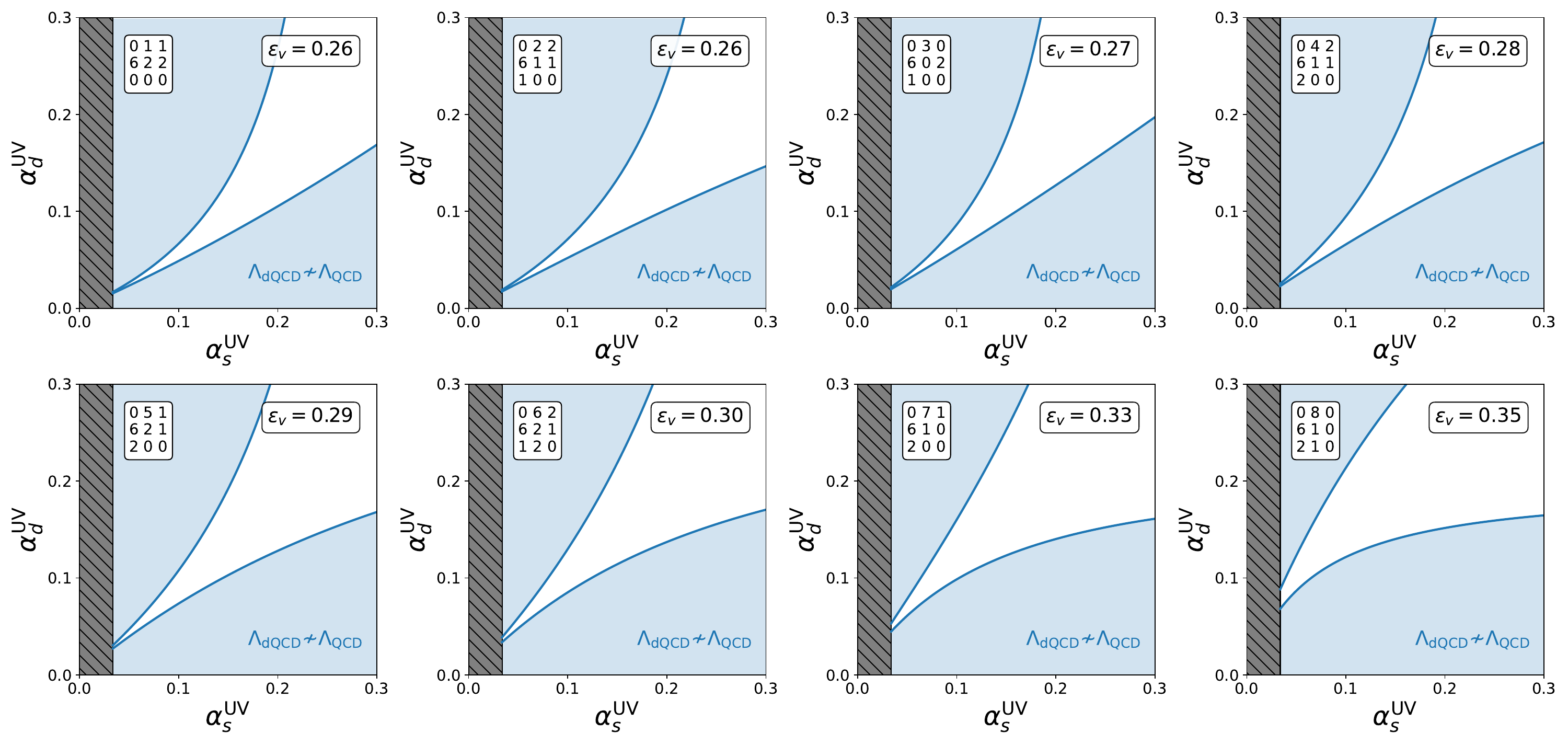}
    \caption{The UV coupling parameter space for the model with the best $\epsilon_v$ for each $n_{dq}$ in Fig.~\ref{fig:SUSY_Nd3_epsv_ndq.pdf}. The blue contours and grey hatched region are the same as in Fig.~\ref{fig:lowM_lowb_models}. The field content of each model is specified in the same way as in Fig.~\ref{fig:changing_ndq.pdf}.}
    \label{fig:SUSY_Nd3_max_epsv.pdf}
\end{figure*}

In Fig.~\ref{fig:SUSY_Nd3_epsv_ndq.pdf} we also note some interesting trends; in particular, as $n_{dq}$ increases, the maximum possible value of $\epsilon_v$ also increases. To explore this, in Fig.~\ref{fig:SUSY_Nd3_max_epsv.pdf} we show the viable region of the UV coupling parameter space for the model with the best $\epsilon_v$ for each $n_{dq}$. We see that the viable region changes in shape as we increase $n_{dq}$, and it seems to be this change in shape that allows for larger $n_{dq}$ to have larger viability fractions.

In Fig.~\ref{fig:SUSY_Nd3_max_epsv.pdf} the field content of each model is given by the matrix in the top-left corner, using the notation explained in the caption of Fig.~\ref{fig:changing_ndq.pdf}. We note that as $n_{dq}$ increases, the models change from having more fields above the diagonal of the matrix to having more fields below the diagonal. Above the diagonal, the fields are `mostly dark' -- that is, $\mathbf{R}_d$ is higher-dimensional than $\mathbf{R}_v$ -- and below the diagonal they are `mostly visible'. So, it seems that for a given model to reach its maximum value of $\epsilon_v$ with a high (low) number of light dark quarks, it needs more (less) visible field content than dark field content.

To understand this feature, consider $\Lambda_\mathrm{dQCD}$ at $(\alpha_s^\mathrm{UV}, \alpha_d^\mathrm{UV}) = (0.3, 0.3)$. As discussed earlier, for $\epsilon_v$ to be as large as possible we must have $\Lambda_\mathrm{dQCD} \sim \Lambda_\mathrm{QCD}$ at this point. Now consider a model with $n_{dq}= 6$. In such a model, since $N_d = 3$, the running of $\alpha_s$ and $\alpha_d$ will be the same below $M$. So, for visible and dark QCD to have similar confinement scales, the running of $\alpha_s$ and $\alpha_d$ in the full theory must also be similar. The model thus requires similar numbers of mostly-visible and mostly-dark fields; indeed, we see that the model with $n_{dq} = 6$ in Fig.~\ref{fig:SUSY_Nd3_max_epsv.pdf} has equal numbers of mostly-visible and mostly-dark fields.

For $n_{dq} > 6$, the running of $\alpha_d$ is weaker than $\alpha_s$ below $M$. So, to have similar confinement scales for $(\alpha_s^\mathrm{UV}, \alpha_d^\mathrm{UV}) = (0.3, 0.3)$, the running of $\alpha_d$ must also be weaker than $\alpha_s$ in the full theory; for a theory with an infrared ZCFP, this requires that the model has more mostly-visible field content than mostly-dark field content. For $n_{dq} < 6$, the opposite reasoning applies.

This link between $n_{dq}$ and the selection of field content also explains the distribution we see in the number of models for each $n_{dq}$, which peaks at $n_{dq} = 6$. This is the number of light dark quarks for which models with large $\epsilon_v$ should have similar numbers of mostly-visible and mostly-dark fields; in the set of models we consider, most models will have such a selection of field content.

We also note that none of the models in Fig.~\ref{fig:SUSY_Nd3_max_epsv.pdf} contain a $\rep{6}{6}$ field. This is interesting, and seems to occur because adding a jointly-charged higher-dimensional representation greatly increases the $\beta$-function coefficients. Larger $\beta$-function coefficients cause the contours to shrink together slightly, which decreases the maximum possible value of $\epsilon_v$. This can be clearly seen when comparing the model with $n_{dq} = 6$ in Fig.~\ref{fig:SUSY_Nd3_max_epsv.pdf} (which has $\epsilon_v = 0.3$) to those in Fig.~\ref{fig:lowM_lowb_models}. These models, which have low values of $b_v^{(0)}$, have $\epsilon_v \sim 0.34$, and $\epsilon_v$ decreases slightly as $b_v^{(0)}$ increases from left to right.

This suggests that, to obtain the largest values of $\epsilon_v$, we in fact do not want to include too many fields in higher-dimensional representations of visible and dark QCD. In our set of models, for the 243 models with a single $\rep{6}{6}$ field, only 2 ($\sim1\%$) have $\epsilon_v > 0.3$, and there are no models with two $\rep{6}{6}$ fields and $\epsilon_v > 0.3$. However, this only affects the models with the very largest values of $\epsilon_v$, and the proportion of models with $\epsilon_v > 0.25$ does not change much when $\rep{6}{6}$ fields are included: for $n^{\rep{6}{6}} = 1~(2)$, 37\% (41\%) of models have $\epsilon_v > 0.25$.

Another effect of adding in $\rep{6}{6}$ fields is to decrease the possible range of $n_{dq}$ such that $\epsilon_v$ is a maximum. For $n^{\rep{6}{6}} = 1$, $n_{dq}$ can only be between 3 and 7, and for $n^{\rep{6}{6}} = 2$, $n_{dq}$ can only be between 4 and 7. So, in this sense, including a $\rep{6}{6}$ or similarly higher-dimensional field makes the required number of light dark quarks more predictive.

The last two features of the results in Fig.~\ref{fig:SUSY_Nd3_epsv_ndq.pdf} that we'd like to explain are the cluster of models with $n_{dq} = 1$ and very small viability fractions, and the decrease in the minimum possible value of $\epsilon_v$ as $n_{dq}$ increases.

To explain the first of these, we note that the models with small $\epsilon_v$ all have many more mostly-dark fields than mostly-visible fields. As we previously showed, such models require a small value of $n_{dq}$ to obtain a decently large $\epsilon_v$. However, for these particular models the running of $\alpha_d$ in the full theory is so much stronger than $\alpha_s$ that even for $n_{dq} = 1$ the running of $\alpha_d$ below $M$ is not strong enough to obtain $\Lambda_\mathrm{dQCD} \sim \Lambda_\mathrm{QCD}$ at $(\alpha_s^\mathrm{UV}, \alpha_d^\mathrm{UV}) = (0.3, 0.3)$. The viable region for these models will thus look like the right-most plot of Fig.~\ref{fig:changing_ndq.pdf}; increasing $n_{dq}$ will only `rotate' the viable region further anti-clockwise, making $\epsilon_v$ even smaller.

To understand the second of these features, we again consider the value of the dark confinement scale at $(\alpha_s^\mathrm{UV}, \alpha_d^\mathrm{UV}) = (0.3, 0.3)$; since $\epsilon_v$ reaches its maximum value when $\Lambda_\mathrm{dQCD}\sim1$~GeV at that point, then the minimum value of $\epsilon_v$ occurs when $\Lambda_\mathrm{dQCD}$ is as far from 1~GeV as possible at that point. As we increase $n_{dq}$ for a given set of heavy field content, $\Lambda_\mathrm{dQCD}$ decreases until for some value of $n_{dq}$ it takes a value below 1 GeV. We now recall that $\Lambda_\mathrm{dQCD}$ changes by a greater amount when we add a light dark quark to a model that already has a large value of $n_{dq}$ (see Eqn.~\ref{eqn:changing_ndq}); so, for larger $n_{dq}$, the dark confinement scale in the top-right corner of parameter space can be further from 1 GeV, leading to a lower minimum value of $\epsilon_v$.

\subsubsection{$N_d = 4$}

With $N_d = 4$ we consider a set of selections of heavy field content with at most two of each new field, where the new fields we introduce are in the $\rep{3}{4}$, $\rep{6}{1}$, $\rep{1}{6}$, $\rep{6}{4}$, $\rep{3}{6}$, and $\rep{6}{6}$ representations of $SU(3)_\mathrm{QCD}\times SU(4)_\mathrm{dQCD}$. This is an equivalent set of field content selections to those with $N_d = 3$, as in both cases we only include fields in the fundamental and lowest-dimensional non-fundamental representations of $SU(N_d)_\mathrm{dQCD}$. 

\begin{figure}
    \centering
    \includegraphics[width=0.485\textwidth]{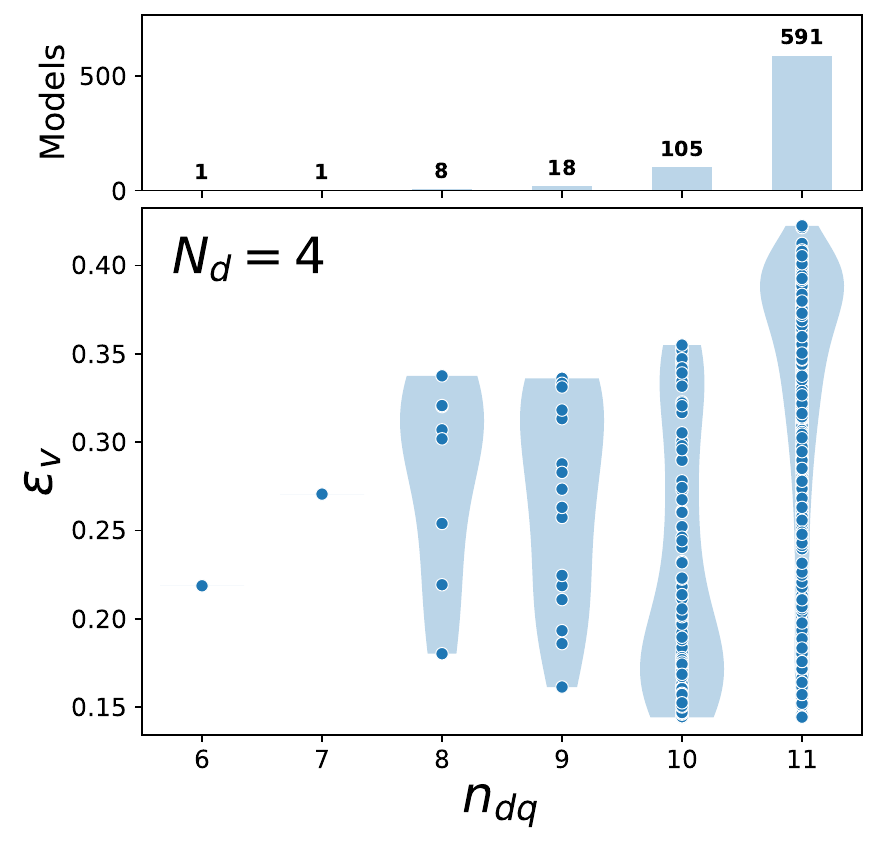}
    \caption{The features of this plot are the same as Fig.~\ref{fig:SUSY_Nd3_epsv_ndq.pdf}, but for a set of 724 models with $N_d = 4$. Each model contains at most two of each new heavy field, where the fields introduced are in the $\rep{3}{4}$, $\rep{6}{1}$, $\rep{1}{6}$, $\rep{6}{4}$, $\rep{3}{6}$, and $\rep{6}{6}$ representations of $SU(3)_\mathrm{QCD}\times SU(4)_\mathrm{dQCD}$.}
    \label{fig:SUSY_Nd4_epsv_ndq.pdf}
\end{figure}

There are 729 such selections of field content; in this case there are only 5 models that lack an infrared ZCFP. As before, for each choice of new heavy field content we find $n_{dq}$ such that $\epsilon_v$ is a maximum, and in Fig.~\ref{fig:SUSY_Nd4_epsv_ndq.pdf} we plot the maximum values of $\epsilon_v$ against $n_{dq}$ for this set of 724 models. 

The results here seem very promising. The best value of $\epsilon_v$ is now $\sim$0.42, an improvement on $N_d = 3$. In addition, while $\epsilon_v$ can be as low as $0.15$ for some models, the majority of models have a decent viability fraction, with 65\% of models having $\epsilon_v > 0.3$. So, with $N_d = 4$, it appears that models with an infrared ZCFP are generically viable.

We also note that the number of light dark quarks required for $\epsilon_v$ to be a maximum is generally larger, with $n_{dq}$ up to 11; in fact, we see that the majority of models ($\sim$80\%) have $n_{dq} = 11$. To understand this behaviour, we first note that to have $\epsilon_v > 0$ requires $n_{dq} \leq 12$. This occurs for the same reasoning we gave earlier; that is, for $n_{dq} > 12$, $\Lambda_\mathrm{dQCD}^\mathrm{max} < 0.2$~GeV. However, for $N_d = 3$ we required $n_{dq} \leq 9$; the fact that larger $n_{dq}$ is allowed when $N_d = 4$ is due to the larger negative gluonic contribution to the coefficients of $\beta_d$. Additionally, for $n_{dq} = 12$ we have $\Lambda_\mathrm{dQCD}^\mathrm{max} < 5$~GeV. This means that $\epsilon_v$ can only be very small, despite being non-zero. For this reason, we see that there are no models for which the maximum $\epsilon_v$ occurs for $n_{dq} = 12$.

So, having shown that $n_{dq} \leq 11$ is necessary for $\epsilon_v$ to be a maximum for a given selection of heavy field content, we now explain why the majority of models have exactly this value. We first recall that for $N_d = 3$, most models have $n_{dq} = 6$ as well as similar numbers of mostly-visible and mostly-dark fields. This is because the running of $\alpha_s$ and $\alpha_d$ is then similar both above and below $M$, such that visible and dark QCD have similar confinement scales for $(\alpha_s^\mathrm{UV}, \alpha_d^\mathrm{UV}) = (0.3, 0.3)$, which ensures that $\epsilon_v$ is at a maximum. For $N_d = 4$, again due to the larger negative gluonic contribution to $\beta_d$, models with similar numbers of mostly-visible and mostly-dark fields now have weaker running for $\alpha_d$ than for $\alpha_s$ in the full theory. So, we also need the running of $\alpha_d$ to be weaker than $\alpha_s$ after the new heavy fields decouple, and it is only for $n_{dq} = 11$ that the running of $\alpha_d$ below $M$ in these models is weak enough to have $\Lambda_\mathrm{dQCD} \sim \Lambda_\mathrm{QCD}$ at $(\alpha_s^\mathrm{UV}, \alpha_d^\mathrm{UV}) = (0.3, 0.3)$.

\begin{figure*}
    \centering
    \includegraphics[width=0.8\textwidth]{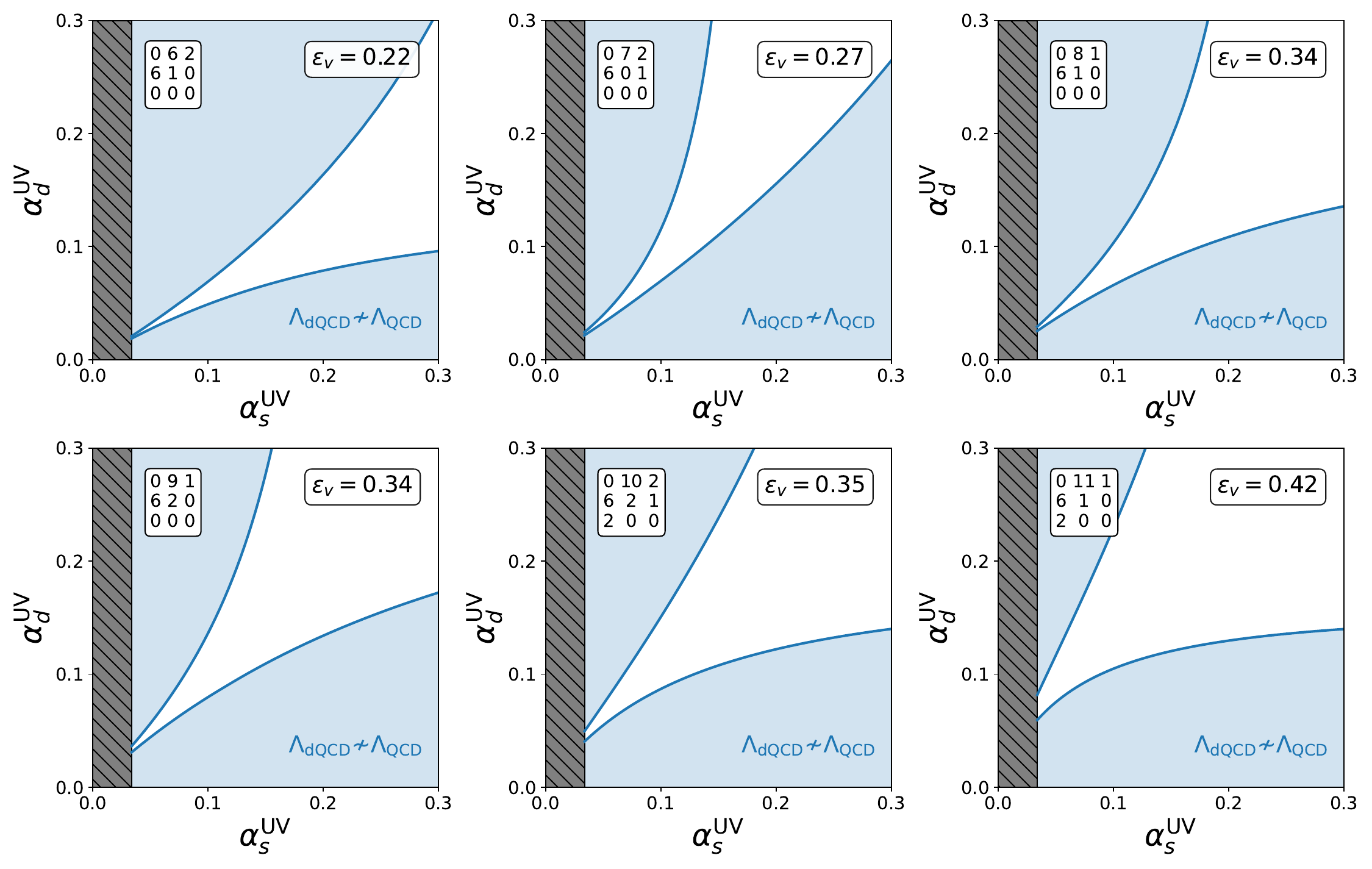}
    \caption{The UV coupling parameter space for the model with the best $\epsilon_v$ for each $n_{dq}$ in Fig.~\ref{fig:SUSY_Nd4_epsv_ndq.pdf}. The blue contours and grey hatched region are the same as in Fig.~\ref{fig:lowM_lowb_models}. The field content of each model is specified in the same way as in Fig.~\ref{fig:changing_ndq.pdf}, but here the columns correspond to $\mathbf{R}_d = \mathbf{1}, \mathbf{4}, \mathbf{6}$ from left to right.}
    \label{fig:SUSY_Nd4_max_epsv.pdf}
\end{figure*}

So, models with similar numbers of mostly-visible and mostly-dark fields -- that is, most models -- require $n_{dq} = 11$. As before, models with more mostly-visible fields require lower $n_{dq}$ to obtain their maximum $\epsilon_v$, as we see in Fig.~\ref{fig:SUSY_Nd4_max_epsv.pdf} where we show the viable region of the UV coupling parameter space for the models with the best $\epsilon_v$ for each $n_{dq}$. Again using our earlier reasoning, models with more mostly-dark fields should require larger $n_{dq}$ to obtain their maximum $\epsilon_v$. However, as $n_{dq}$ cannot be greater than 11, the models with more mostly-dark fields instead just have $n_{dq} = 11$ and smaller values of $\epsilon_v$. We observe this behaviour in Fig.~\ref{fig:SUSY_Nd4_epsv_ndq.pdf}, where for $n_{dq} = 11$ there is a `tail' of models with lower viability fractions below a cluster around $\epsilon_v \sim 0.39$.

We now wish to understand why there is such a cluster of models at high $\epsilon_v$ for $n_{dq} = 11$, as it is this feature that gives us a large proportion of viable models with $N_d = 4$. Firstly, the fact that such values of $\epsilon_v$ are even possible is due to the fact that these models have a large number of light dark quarks. As we see in Fig.~\ref{fig:SUSY_Nd4_max_epsv.pdf}, the characteristic shape of the viable regions changes with $n_{dq}$, and the maximum possible value of $\epsilon_v$ increases accordingly. Secondly, to explain the large number of models with $\epsilon_v \sim 0.39$, consider a simple model with $n_{dq} = 11$ and a single heavy $\rep{6}{6}$ field. It happens that this model has $\epsilon_v = 0.39$. We now consider adding in `equally visible and dark' field content to this model: for example, an extra $\rep{6}{6}$ field, or a $\rep{6}{1}$ and a $\rep{1}{6}$ field. This does not substantially change the relative strength of running for $\alpha_s$ and $\alpha_d$ in the full theory, and so the value of $\epsilon_v$ does not change much in these models; indeed, the two example models mentioned in the previous sentence both have $\epsilon_v = 0.39$. So, all such models we construct in this way will have $\epsilon_v \sim 0.39$, which leads to the observed cluster of models in Fig.~\ref{fig:SUSY_Nd4_epsv_ndq.pdf}.

We note that for $N_d = 3$ the results feature a similar cluster of models. However, in this case, the model with $n_{dq} = 6$ and a single heavy $\rep{6}{6}$ field only has $\epsilon_v = 0.21$. So for $N_d = 3$, the cluster of models is at $\epsilon_v \sim 0.21$ for $n_{dq} = 6$, as we observe in Fig.~\ref{fig:SUSY_Nd3_epsv_ndq.pdf}.

Finally, as before, we see that none of the models in Fig.~\ref{fig:SUSY_Nd4_max_epsv.pdf} contain a $\rep{6}{6}$ field, since adding a jointly-charged higher-dimensional representation decreases the maximum possible value of $\epsilon_v$. However, in this case including $\rep{6}{6}$ fields in a model does not reduce the proportion of models with decent $\epsilon_v$. In fact, for $n^{\rep{6}{6}} = 2$ it substantially increases the proportion of models, with 75\% of those models having $\epsilon_v > 0.3$.

So, the conclusion of this subsection is that there is a simple set of requirements that generically lead us to viable models: that is, $N_d = 4$, $n_{dq} = 11$, and similar numbers of mostly-visible and mostly-dark fields.

\subsubsection{$N_d = 2$}

\begin{figure}
    \centering
    \includegraphics[width=0.485\textwidth]{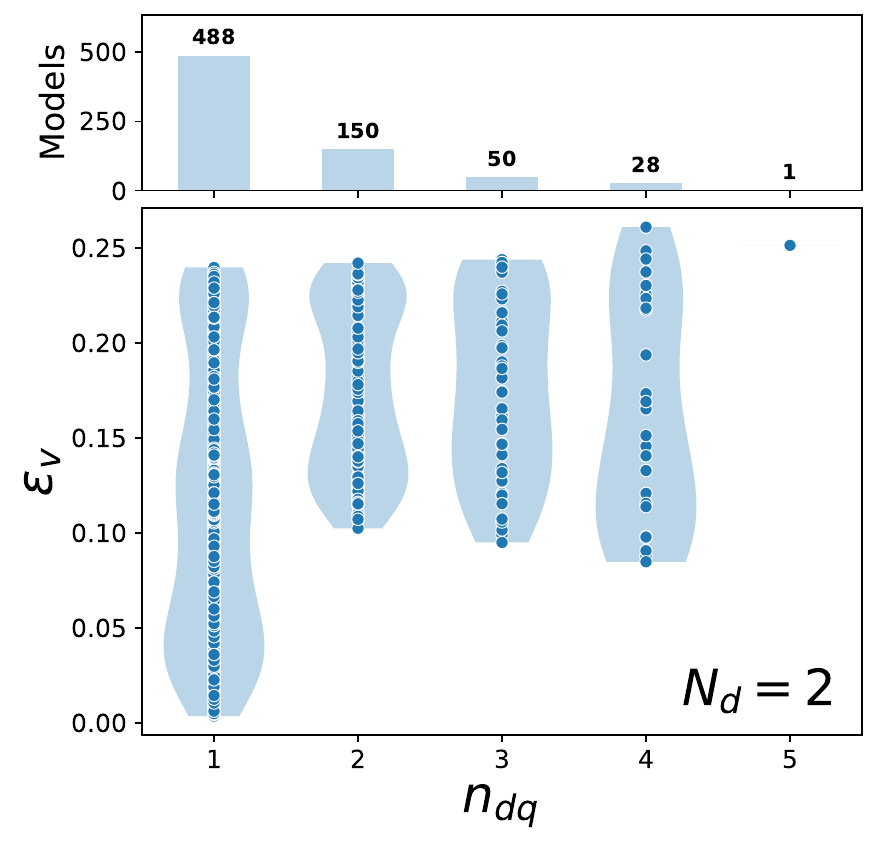}
    \caption{The features of this plot are the same as Fig.~\ref{fig:SUSY_Nd3_epsv_ndq.pdf}, but for a set of 717 models with $N_d = 2$. Each model contains at most two of each new heavy field, where the fields introduced are in the $\rep{3}{2}$, $\rep{6}{1}$, $\rep{1}{3}$, $\rep{6}{2}$, $\rep{3}{3}$, and $\rep{6}{3}$ representations of $SU(3)_\mathrm{QCD}\times SU(2)_\mathrm{dQCD}$.}
    \label{fig:SUSY_Nd2_epsv_ndq.pdf}
\end{figure}

Finally in this section, for completeness we briefly mention the case with $N_d = 2$. With $N_d = 2$ we consider a set of selections of heavy field content with at most two of each new field, where the new fields we introduce are in the $\rep{3}{2}$, $\rep{6}{1}$, $\rep{1}{3}$, $\rep{6}{2}$, $\rep{3}{3}$, and $\rep{6}{3}$ representations of $SU(3)_\mathrm{QCD}\times SU(2)_\mathrm{dQCD}$. Again, this is an equivalent set of field content selections to before. Of the 729 selections of field content, there are 12 models that lack an infrared ZCFP. Again, for each choice of new heavy field content we find $n_{dq}$ such that $\epsilon_v$ is a maximum. In Fig.~\ref{fig:SUSY_Nd2_epsv_ndq.pdf} we plot the maximum values of $\epsilon_v$ against $n_{dq}$ for this set of 717 models. 

The results in this case are worse than both $N_d = 3$ and $N_d = 4$. When $N_d = 4$, the large negative gluonic contributions to $\beta_d$ mean that $n_{dq}$ must take larger values; here, the gluonic contributions are smaller, and so $n_{dq}$ takes small values, with the majority of models having $n_{dq} = 1$. The maximum possible value of $\epsilon_v$ is smaller for models with smaller values of $n_{dq}$, and so the best value of $\epsilon_v$ is just above 0.25. So, only a very small proportion of models have a decently large viability fraction, with only two models having $\epsilon_v > 0.25$.

\section{Embedding a viable model within a complete asymmetric dark matter model}\label{sec:adm}

The viable models we identified in the previous section predict similar masses for visible and dark matter, $m_\mathrm{DM} \sim m_p$. To fully explain the cosmological coincidence with one of these models, we now turn to the question of obtaining similar number densities, $n_\mathrm{DM} \sim n_B$, by incorporating some asymmetric dark matter mechanism.

Bai and Schwaller~\cite{Bai:2013xga} considered a simple leptogenesis-like idea that took advantage of the fields charged under both visible and dark QCD. In their model, heavy right-handed neutrinos (RHNs) $N_i$ generate an asymmetry in a new jointly-charged field through their $CP$-violating, out-of-equilibrium decays. The subsequent decays of this jointly-charged field distribute the asymmetry to the visible baryons and dark baryons in rougly equal amounts, naturally obtaining related number densities.

We note, however, that the $N_i$ can also couple to visible leptons. In the absence of any new physics that forbids such a coupling, $N_i$ decays thus also introduce an initial visible lepton asymmetry; following sphaleron reprocessing, this asymmetry will be partially converted into a baryon asymmetry, just as in standard thermal leptogenesis~\cite{Fukugita:1986hr}. This weakens the relationship between the final visible and dark baryon asymmetries, as it now depends on the couplings of $N_i$ to the visible leptons and the jointly-charged fields; this is similar to the ADM models of Refs.~\cite{An:2009vq,Falkowski:2011xh}. In this section we first consider this leptogenesis mechanism in a (non-supersymmetric) viable model with a low mass scale $M$, as in Sec.~\ref{sec:lowM_models}. We find that, assuming similar couplings of $N_i$ to the visible leptons and jointly-charged fields, we still obtain $n_B/n_D \sim \mathcal{O}(1)$.

We then give a version of this leptogenesis mechanism in a supersymmetric viable model with $N_d = 4$. We also address the additional difficulties introduced by supersymmetry, where (i) thermal leptogenesis now faces the gravitino problem~\cite{Kawasaki:2008qe} and (ii) to have the dark baryon be the dark matter candidate, we must ensure that the lightest supersymmetric partner (LSP) is either unstable or only comprises a subdominant component of the dark matter density.

Finally we consider the annihilation of the symmetric component of the dark matter relic density.

\subsection{Leptogenesis in a model with a low mass scale}\label{sec:lowM_LG}

In this section we consider a version of the leptogenesis model proposed by Bai and Schwaller, looking to incorporate it within one of the viable models identified in Section~\ref{sec:lowM_models}. In Ref.~\cite{Bai:2013xga}, equal and opposite asymmetries are generated in a $\rep{\bar{3}}{3}_{1/3}$ fermion and a $\rep{\bar{3}}{3}_{1/3}$ scalar through their Yukawa interaction with the RHNs $N_i$. A similar term can produce an asymmetry in any pair of fermion and scalar fields with the same quantum numbers under $SU(3)_\mathrm{QCD} \times SU(N_d)_\mathrm{dQCD} \times U(1)_Y$. The models in Table~\ref{tab:lowM_nofermion_models} do not feature such a term, as they only contain new fermions. Of the models in Table~\ref{tab:lowM_maxtriplet_models}, all but one feature a $\rep{3}{3}$ fermion and a $\rep{3}{3}$ scalar; the following simple model can be embedded in any of these four models.

Consider a model with a $\rep{3}{3}_{-1/3}$ Dirac fermion $\Psi_1$ and a $\rep{3}{3}_{-1/3}$ complex scalar $\Phi$. The relevant Lagrangian terms involving the three RHNs $N_i$ are
\begin{equation}\label{eqn:LG_Ni_interactions}
    -\mathcal{L} \supset \frac{1}{2}M_{N_i}N_i^2 + k_iN_i\bar{\Psi}_1\Phi + \lambda_{i\alpha}N_iL_\alpha H + \text{h.c.},
\end{equation}
where we work in a basis where the neutrino masses are real and diagonal. The $CP$-violating decays of the $N_i$ produce asymmetries in the new jointly-charged fields, $\Delta n_{\Psi_1} \equiv n_{\Psi_1} - n_{\bar{\Psi}_1}$ and $\Delta n_{\Phi} = -\Delta n_{\Psi_1}$, in addition to an asymmetry in visible leptons, $\Delta n_{L}$. In the hierarchical limit $M_{N_1} < M_{N_2} < M_{N_3}$, the $CP$-asymmetries from each decay are~\cite{Falkowski:2011xh}
\begin{align}
    \epsilon_L &\simeq \frac{3}{8\pi}\frac{\text{Im}\left[\left(\left[\lambda\lambda^\dagger\right]_{21} + 3k_2k_1^\star\right)\left[\lambda\lambda^\dagger\right]_{21}\right]}{2\left[\lambda\lambda^\dagger\right]_{11} + 9|k_1|^2}\frac{M_{N_1}}{M_{N_2}},\\
    \epsilon_{\Psi_1} &\simeq \frac{9}{8\pi}\frac{\text{Im}\left[\left(\left[\lambda\lambda^\dagger\right]_{21} + 5k_2k_1^\star\right)k_2k_1^\star\right]}{2\left[\lambda\lambda^\dagger\right]_{11} + 9|k_1|^2}\frac{M_{N_1}}{M_{N_2}}.
\end{align}

So, the ratio between the two asymmetry parameters depends upon the relative sizes of the Yukawa couplings, and can be sizeable if there is a hierarchy between the couplings. If the $N_i$ are responsible for the small masses of the SM neutrinos through the seesaw mechanism, then the Yukawa couplings $\lambda_{i\alpha}$ may be hierarchical to generate the observed pattern of masses. Assuming that all other Yukawa couplings $k_i$ are $\mathcal{O}(1)$, then the $CP$-asymmetry in visible leptons will be much smaller than the jointly-charged fields. If we assume more generically that all Yukawa couplings are of the same order, then $\epsilon_L/\epsilon_{\Psi_1} \sim 2/9$, where the larger asymmetry in $\Psi_1$ is due to its higher dimension.

The final asymmetries for $\Psi_1$, $\Phi$, and $L$ are altered by washout effects. For each decay, the washout regime in which it occurs depends upon the decay parameter
\begin{equation}
    K_x \equiv \frac{\Gamma_{N_1}\text{Br}_x}{H(M_{N_1})}, \quad x = \Psi_1, L,
\end{equation}
where $\Gamma_{N_1} = (2\left[\lambda\lambda^\dagger\right]_{11} + 9|k_1|^2)M_{N_1}/16\pi$ is the total decay rate of $N_1$, $\text{Br}_x$ is the branching ratio of its decay into $x$, and $H(M_{N_1})$ is the Hubble expansion rate at $T = M_{N_1}$. As in Ref.~\cite{Bai:2013xga}, with $M_{N_1} \sim 10^{13}$~GeV and $\mathcal{O}(1)$ Yukawa couplings, both decays are in the strong washout regime where $K \gtrsim 1$. Then, the final asymmetry can be estimated by~\cite{Buchmuller:2004nz}
\begin{equation}
    Y_{\Delta x}^\infty = \frac{2}{z(K_x)K_x}\epsilon_x Y_{N_1}^\mathrm{eq}(0),
\end{equation}
where the number densities are normalised by entropy density as abundance yields $Y_x = n_x/s$ and $Y_{\Delta x} = Y_x - Y_{\bar{x}}$, $z(K_x)$ is the washout decoupling temperature, and $Y_{N_i}^\mathrm{eq}(0) \sim 0.4/g_\star$ is the initial equilibrium abundance of $N_1$ with $g_\star \sim 300$ the number of relativistic degrees of freedom at $T \sim M_{N_1}$. In Ref.~\cite{Bai:2013xga} this asymmetry was found to be large enough to generate the observed visible and dark baryon asymmetries with a mild hierarchy in the RHN masses, $K = 10\text{---}100$, corresponding to $z(K) = 4\text{---}7$.

The ratio between the asymmetries generated by the decay of $N_i$ is
\begin{equation}
    \frac{\Delta n_L}{\Delta n_{\Psi_1}} = \frac{\epsilon_L\text{Br}_{\Psi_1}z(K_{\Psi_1})}{\epsilon_{\Psi_1}\text{Br}_Lz(K_L)}.
\end{equation}

In the case where all Yukawa couplings are of a similar size, this simplifies to $\Delta n_L/\Delta n_{\Psi_1} \sim z(K_{\Psi_1})/z(K_L) = 1.3\text{---}1.5$ for $K = 10\text{---} 100$.

We now consider the distribution of these asymmetries into visible and dark baryons through the further decays of $\Psi_1$ and $\Phi$. This occurs just as in Ref.~\cite{Bai:2013xga}; we introduce a second $\rep{3}{3}_{2/3}$ fermion $\Psi_2$ and dark quarks $X \sim \rep{1}{3}_0$, with interactions
\begin{equation}\label{eqn:LG_decay_interactions}
    -\mathcal{L} \supset \kappa_1\bar{\Psi}^c_1\bar{\Psi}_2\Phi + \kappa_2\bar{\Psi}_2\Phi e_R^c + \kappa_3\Phi\bar{d}_RX^c  + \text{h.c.}
\end{equation}
As in Ref.~\cite{Bai:2013xga}, the interactions of Eqs.~\ref{eqn:LG_Ni_interactions} and \ref{eqn:LG_decay_interactions} admit an accidental $\mathbb{Z}_2$ symmetry under which $X$, $\Phi$, $\Psi_1$, and $e_R$ are odd. This ensures the stability of the dark baryon, which is a confined state of three light dark quarks $X$.

Assuming a slight mass hierarchy $M_{\Psi_1} > M_{\Psi_2} > M_{\Phi}$, the heavy fields decay via $\Psi_1 \rightarrow \bar{\Psi}_2 + \Phi^\dagger$ followed by $\Psi_2 \rightarrow \Phi^\dagger + \bar{e}_R$ and $\Phi \rightarrow d_R + X$. This results in baryon, lepton, and dark baryon number densities given by
\begin{align}
    &n_B = -\Delta n_{\Psi_1},\\
    &n_L = \Delta n_{\Psi_1} + \Delta n_{L},\\
    &n_D = -\Delta n_{\Psi_1}.
\end{align}

Then, following sphaleron reprocessing, the standard leptogenesis relation $n_B = 28/79n_{B-L}$ gives
\begin{equation}
    \frac{n_B}{n_D} = \frac{28}{79}\left(2 + \frac{\Delta n_{L}}{\Delta n_{\Psi_1}}\right).
\end{equation}

As discussed earlier, $\Delta n_{L}/\Delta n_{\Psi_1} \ll 1$ may occur if there is a hierarchy amongst the Yukawa couplings $\lambda_{i\alpha}$ of $N_i$ to the visible leptons, but the $k_i$ are all $\mathcal{O}(1)$. If all Yukawa couplings are similar, then following strong washout we have $\Delta n_{L}/\Delta n_{\Psi_1} \sim 1.4$. In both of these cases the ratio of number densities is $\mathcal{O}(1)$,
\begin{equation}
\frac{n_B}{n_D} \sim 
\begin{cases}
    0.7 & \Delta n_{L} \ll \Delta n_{\Psi_1},\\
    1.2 & \mathcal{O}(1)\text{ Yukawa couplings}.
\end{cases}
\end{equation}

The number densities would only be unrelated in the unlikely case that there is a strong coupling hierarchy $k_i \ll \lambda_{i\alpha}$. So, this model manages to relate the number densities of visible and dark matter, despite the multiple decay pathways of the RHNs $N_i$.

\subsection{Supersymmetric leptogenesis}\label{sec:SUSY_LG}

We now consider implementing this leptogenesis idea in a viable supersymmetric model. We work with $N_d = 4$, since in Section~\ref{sec:SUSY_models} we found this allows us to generically obtain viable models. Thermal leptogenesis in supersymmetry has been explored in detail~\cite{Covi:1996wh, Giudice:2003jh, Buchmuller:2005eh, Fong:2010qh}, and proceeds in a fairly similar fashion to the standard case. We also note that there are some existing supersymmetric ADM models with leptogenesis-like mechanisms~\cite{Haba:2010bm, Choi:2012ba}, but their implementations differ from ours.

Before giving the details of our model, we mention that within supersymmetry there is a natural mechanism for generating particle asymmetries, the Affleck-Dine mechanism~\cite{Affleck:1984fy, Linde:1985gh, Dine:1995kz, Dine:2003ax}, using oscillations along the naturally-occurring flat directions of the superpotential. This is often implemented in ADM scenarios~\cite{Bell:2011tn, Cheung:2011if, vonHarling:2012yn}, and could be used here to generate an initial asymmetry in a jointly-charged field which subsequently decays into visible and dark baryons. This is a nice idea, but we do not consider it further here.

\begin{table}[]
\begin{tabular}{c|cccc}
         & $SU(3)_\mathrm{QCD} \times SU(4)_\mathrm{dQCD}$ & $Y$     & $\tilde{B}$ & $\tilde{L}$ \\ \hline
$N_i$    & $(\mathbf{1}, \mathbf{1})$                      & $0$      & $0$         & $-1$        \\
$X_i$    & $(\mathbf{1}, \mathbf{4})$                      & $0$      & $-1/4$      & $1/4$       \\
$\Phi_1$ & $(\mathbf{3}, \mathbf{6})$                      & $2/3$    & $-1/6$      & $1/2$       \\
$\Phi_2$ & $(\mathbf{3}, \mathbf{6})$                      & $2/3$    & $-1/6$      & $-1/2$      \\
$\Phi_3$ & $(\mathbf{3}, \mathbf{4})$                      & $2/3$    & $1/12$      & $1/4$       \\
$\Phi_4$ & $(\mathbf{3}, \mathbf{4})$                      & $-1/3$   & $1/12$      & $1/4$   
\end{tabular}
\caption{The field content of our supersymmetric leptogenesis model. Note that all fields except the $N_i$ are Dirac; that is, for each chiral supermultiplet $\Phi \sim \RvRd_y$ with $\tilde{B} = x$ and $\tilde{L} = y$ we also introduce another chiral supermultiplet $\bar{\Phi} \sim \rep{\bar{\mathbf{R}}_v}{\bar{\mathbf{R}}_d}_{-y}$ with $\tilde{B} = -x$ and $\tilde{L} = -y$, allowing for the mass term $M\bar{\Phi}\Phi$.}
\label{tab:SUSY_LG_model}
\end{table}

The new field content required for this model is given in Table~\ref{tab:SUSY_LG_model}, where we specify the gauge charges under $SU(3)_\mathrm{QCD} \times SU(4)_\mathrm{dQCD} \times U(1)_Y$, along with two other charge assignments that we will discuss shortly. Here, the $N_i$ are singlet chiral superfields with fermionic neutrino and scalar sneutrino components, and we refer to the fermionic and scalar components of the light fields $X_i$ as the dark quarks and dark squarks, respectively. The model is an extension of the $R$-parity--conserving Minimal Supersymmetric Standard Model (MSSM) with superpotential
\begin{align}
    &W = W_\mathrm{MSSM} + W_\mathrm{mass} + W_\mathrm{asym} + W_\mathrm{decay}, \\
    &W_\mathrm{mass} = \frac{1}{2}M_{N_i}N_i^2   +   m_{X_i} \bar{X}_i X_i   +   M_{\Phi_i} \bar{\Phi}_i \Phi_i, \label{eq:W_mass} \\
    &W_\mathrm{asym} = \lambda_{i\alpha} N_i L_a H_u   +   k_i N_i \Phi_1 \bar{\Phi}_2, \label{eq:W_asym} \\
    &W_\mathrm{decay} = \kappa_1 \bar{\Phi}_1 \Phi_3 X   +   \bar{\kappa}_1 \Phi_1 \bar{\Phi}_3 \bar{X}   +   \kappa_2 \Phi_2 \Phi_4^2 \nonumber \\ 
    & \hspace{2cm} + \bar{\kappa}_2 \bar{\Phi}_2 \bar{\Phi}_4^2   +   \kappa_3 \Phi_3 \bar{u} \bar{X}   +   \kappa_4 \Phi_4 \bar{d} \bar{X}, \label{eq:W_decay}
\end{align}
where we split the additional terms in the superpotential into the mass terms for the new fields, the interaction terms responsible for the initial asymmetry generation through the decay of the $N_i$, and the interaction terms responsible for the subsequent distribution of these asymmetries into visible and dark baryon asymmetries. For $W_\mathrm{decay}$ we suppress the generational indices for the light dark (s)quarks $X_i$, and we work in a basis where the masses of the $X_i$ and the $N_i$ are real and diagonal. Note also that we allow the masses of the new heavy fields $\Phi_i$ to differ from each other slightly, but they all lie close to the required mass scale of new physics, $M$.

The terms in the superpotential (bar the Majorana masses of the $N_i$) feature two $U(1)$ symmetries, which we treat as extensions of standard baryon and lepton number, naming them $\tilde{B}$ and $\tilde{L}$ respectively\footnote{These charges can also be used to define a discrete `matter parity' symmetry, $P_{\tilde{M}} = (-1)^{3\tilde{B} - \tilde{L}}$. This gives all the fields of the MSSM their standard matter parity assignments, and so can be considered as an extension of the standard matter- or $R$-parity. The superpotential of this model respects this symmetry.}; their assignments are given in the last two columns of Table~\ref{tab:SUSY_LG_model}. The masses of the $N_i$ softly break the extended lepton number, and thus allow for an asymmetry in $\tilde{L}$ to be generated through the $CP$-violating, out-of-equilibrium decays implied by the superpotential terms of Eqn.~\ref{eq:W_asym}. If $M_{N_i} > M$, then these decays of the heavy neutrinos and sneutrinos produce asymmetries in the fermionic and scalar components of the new heavy fields $\Phi_1$ and $\Phi_2$, where the total asymmetries in each field satisfy $\Delta n_{\Phi_1} = -\Delta n_{\Phi_2}$; this is in addition to an asymmetry in visible leptons $\Delta n_L$. 

The further decays of the new heavy fields then distribute their asymmetries into visible and dark (s)quarks, where the scalar and fermionic parts of each asymmetry equilibrate through rapid gluino-mediated processes. The visible and dark quarks will eventually confine into the visible baryons and the dark baryons that serve as our dark matter candidate. Before considering the sizes of these asymmetries, let us first discuss the stability of the dark baryon $\chi$.

The dark baryons are confined states of 4 light dark quarks, with mass $m_\mathrm{\chi}$ a few times $\Lambda_\mathrm{dQCD}$. These particles have $\tilde{B} = -1$ and $\tilde{L} = 1$, and so are able to decay into SM particles via e.g.~$\chi \rightarrow \bar{n} + \nu$. However, this decay is mediated by all the new heavy fields in Table~\ref{tab:SUSY_LG_model}, and so is extremely strongly suppressed; this makes the dark baryon very long-lived, and thus sufficiently stable to be the cold dark matter.

In fact, at energies below the mass scale of new physics $M$, all interactions between the dark quarks and SM particles are suppressed by powers of $M$ and are thus out of thermal equilibrium. So, at low energies, the particle number of the light dark quarks $X_i$ will be conserved to a very high degree. Let us define the particle number of $X_i$ as $D = 1/4$, so that the dark baryons $\chi$ have $D = 1$; at low energies, this `dark baryon number' $D$ is approximately conserved.

We now consider the size of the asymmetries generated by the decay of the $N_i$. In the hierarchical limit for RHN masses, the $CP$-asymmetries from each decay in Eqn.~\ref{eq:W_asym} are
\begin{align}
    \epsilon_L &\simeq \frac{3}{4\pi}\frac{\text{Im}\left[\left(\left[\lambda\lambda^\dagger\right]_{21} + 6k_2k_1^\star\right)\left[\lambda\lambda^\dagger\right]_{21}\right]}{\left[\lambda\lambda^\dagger\right]_{11} + 9|k_1|^2}\frac{M_1}{M_2},\\
    \epsilon_{\Phi_1} &\simeq \frac{9}{4\pi}\frac{\text{Im}\left[\left(2\left[\lambda\lambda^\dagger\right]_{21} + 19k_2k_1^\star\right)k_2k_1^\star\right]}{\left[\lambda\lambda^\dagger\right]_{11} + 9|k_1|^2}\frac{M_1}{M_2}.
\end{align}

If all Yukawa couplings are of a similar size, then $\epsilon_L/\epsilon_{\Phi_1} \sim 1/9$. In the non-supersymmetric case, we also considered hierarchical Yukawa couplings $\lambda_{i\alpha}$, as may be needed to generate the observed neutrino masses. In this case, however, the $N_i$ are too heavy to provide sufficiently large neutrino masses with perturbative Yukawa couplings. To do so requires $M_{N_i} \lesssim 10^{15}$~GeV, but to generate asymmetries in the new heavy fields we require $M_{N_i} > M$, where the mass scale of physics is at least $10^{15}$~GeV in these models, if not higher (see Table~\ref{tab:susy_min_M})\footnote{In general, the neutrino masses can be generated by some alternative mechanism. However, if we do want the $N_i$ to generate the small neutrino masses, then they must have $M_{N_i} < M$ and so cannot decay into the new heavy fields. To generate a dark baryon asymmetry, the $N_i$ would need to have $CP$-violating decays to the light dark (s)quarks $X_i$. This would require some of the $X_i$ to have differing $\tilde{L}$ charges.}.

Another consequence of large $M_{N_i}$ is that each decay is in the weak washout regime $K \ll 1$. In this regime, the final asymmetry depends depends strongly on the initial conditions for the abundance of $N_1$. Assuming an initial thermal abundance, the final asymmetry can be estimated by~\cite{Buchmuller:2004nz}
\begin{equation}
    Y_{\Delta x}^\infty = \epsilon_x Y_{N_1}^\mathrm{eq}(0), \quad x = \Phi_1, L.
\end{equation}
To avoid overproducing the baryon and dark baryon asymmetries, we cannot have all Yukawa couplings be $\mathcal{O}(1)$. Achieving the observed baryon-to-entropy ratio of $9\times10^{-11}$~\cite{Fong:2012buy} is possible with only somewhat small Yukawa couplings: if some Yukawa couplings are $\mathcal{O}(10^{-3})$, then with $M_{N_2} = 10M_{N_1}$ we have $Y_{\Delta {\Phi_1}}^\infty \sim 10^{-9}\sin{\delta}$, where $\delta$ represents the $CP$-violating phases in the complex Yukawa couplings.

However, since $M_{N_1}$ is very large, obtaining an initial thermal abundance requires a very high reheating temperature. In the case of vanishing initial $N_1$-abundance, heavy $N_1$ are produced dynamically by inverse decays, leading to a suppressed final asymmetry given by~\cite{Buchmuller:2004nz}
\begin{equation}
    Y_{\Delta x}^\infty \simeq \frac{9\pi^2}{64} K_x^2 \epsilon_x Y_{N_1}^\mathrm{eq}(0), \quad x = \Phi_1, L.
\end{equation}
In this case, assuming that all Yukawa couplings are of a similar order, the final asymmetry in $\Phi_1$ is given by
\begin{multline}
    Y_{\Delta \Phi_1}^\infty \sim 1.4 \times 10^{-8} \left(\frac{k_1}{0.1}\right)^4 \left(\frac{k_2}{0.1}\right)^2 \times\\
    \left(\frac{10^{17}\text{~GeV}}{M_{N_1}}\right)^2 \left(\frac{300}{g_\star}\right) \frac{M_{N_1}}{M_{N_2}} \sin{\delta}.
\end{multline}
and the observed baryon-to-entropy ratio of $9\times10^{-11}$ can be achieved with $\mathcal{O}(1)$ Yukawa couplings and $M_{N_1} = M_{N_2}/10 = 10^{17}$~GeV.

In the weak washout regime, the ratio between the asymmetries generated by the decay of the $N_i$ also depends on its initial conditions, with
\begin{equation}
\frac{\Delta n_L}{\Delta n_{\Phi_1}} = 
\frac{\epsilon_L}{\epsilon_{\Phi_1}} \times\begin{cases}
    1 & \text{thermal $N_1$-abundance,}\\
    \frac{\text{Br}_L^2}{\text{Br}_{\Phi_1}^2} & \text{vanishing $N_1$-abundance}.
\end{cases}
\end{equation}
If all Yukawa couplings are of a similar size, then $\epsilon_L/\epsilon_{\Phi_1} \sim \text{Br}_L/\text{Br}_{\Phi_1} \sim 1/9$. This leads to the initial lepton asymmetry being either somewhat smaller, or much smaller than the initial $\Phi_1$ asymmetry:
\begin{equation}\label{eqn:SUSY_LG_initial_asymmetries}
\frac{\Delta n_L}{\Delta n_{\Phi_1}} \sim 
\begin{cases}
    0.1 & \text{thermal $N_1$-abundance,}\\
    1 \times 10^{-3} & \text{vanishing $N_1$-abundance}.
\end{cases}
\end{equation}

We now consider the distribution of these asymmetries into visible and dark baryons. This occurs through the Yukawa and three-scalar interactions implied by the superpotential terms of Eqn.~\ref{eq:W_decay}. Assuming slight mass hierarchies of $M_{\Phi_1} > M_{\Phi_3}$ and $M_{\Phi_2} > M_{\Phi_4}$, the asymmetries are distributed to baryons and dark baryons via the decay chains $\Phi_1 \rightarrow \Phi_3 + X$ followed by $\Phi_3 \rightarrow u + X$, and $\bar{\Phi}_2 \rightarrow \Phi_4 + \Phi_4$ followed by $\Phi_4 \rightarrow d + X$. Note that we write each decay in terms of the superfields as a shorthand for the various decays of the fermionic and scalar components of each field.

This gives final baryon, lepton, and dark baryon number densities
\begin{align}
    &n_B = \frac{1}{3}(\Delta n_u +\Delta n_d) = \Delta n_{\Phi_1},\\
    &n_L = \Delta n_{L},\\
    &n_D = \frac{1}{4}\Delta n_X = \Delta n_{\Phi_1}.
\end{align}
Note that these asymmetries are consistent with the fact that the decays respect $\tilde{B}$ and $\tilde{L}$ conservation, satisfying $\Delta n_{\tilde{B}} = 0$.

This initial $B-L$ asymmetry is then subject to sphaleron reprocessing. Since the dark quarks are not in thermal equilibrium with the SM bath at the relevant temperatures, they do not take part in the reprocessing, and the $D$ asymmetry is unchanged. In SUSY leptogenesis, the second Higgs doublet needs to be taken into account, resulting in $n_B = 8/23 n_{B-L}$~\cite{Khlebnikov:1996vj}. This gives the final ratio of visible and dark baryon number densities as
\begin{equation}
    \frac{n_B}{n_D} = \frac{8}{23}\left(1 - \frac{\Delta n_{L}}{\Delta n_{\Phi_1}}\right).
\end{equation}

Here, if the initial asymmetries in lepton number and the jointly-charged fields are of a similar size, then there is a cancellation that results in $n_B \ll n_D$. However, if all Yukawa couplings are of a similar size, then $\Delta n_{L} < \Delta n_{\Phi_1}$, where the relative sizes of these asymmetries depend on the initial $N_1$ abundance as in Eqn.~\ref{eqn:SUSY_LG_initial_asymmetries}. The particular ratio of initial asymmetries does not have a strong effect on the final ratio of visible and dark baryon number densities, which will lie between the following $\mathcal{O}(1)$ values:
\begin{equation}
    \frac{n_B}{n_D} \sim 
\begin{cases}
    0.31 & \text{thermal $N_1$-abundance,}\\
    0.35 & \text{vanishing $N_1$-abundance}.
\end{cases}
\end{equation}

To obtain the observed cosmological coincidence of Eqn.~\ref{eqn:cosmo_coincidence} then requires a dark baryon with mass
\begin{equation}
    m_\chi = m_p \frac{\Omega_\mathrm{DM}}{\Omega_\mathrm{VM}}\frac{n_B}{n_D} \sim 1.6\text{---}1.8\text{ GeV}.
\end{equation}

\begin{figure}
    \centering
    \includegraphics[width=0.3\textwidth]{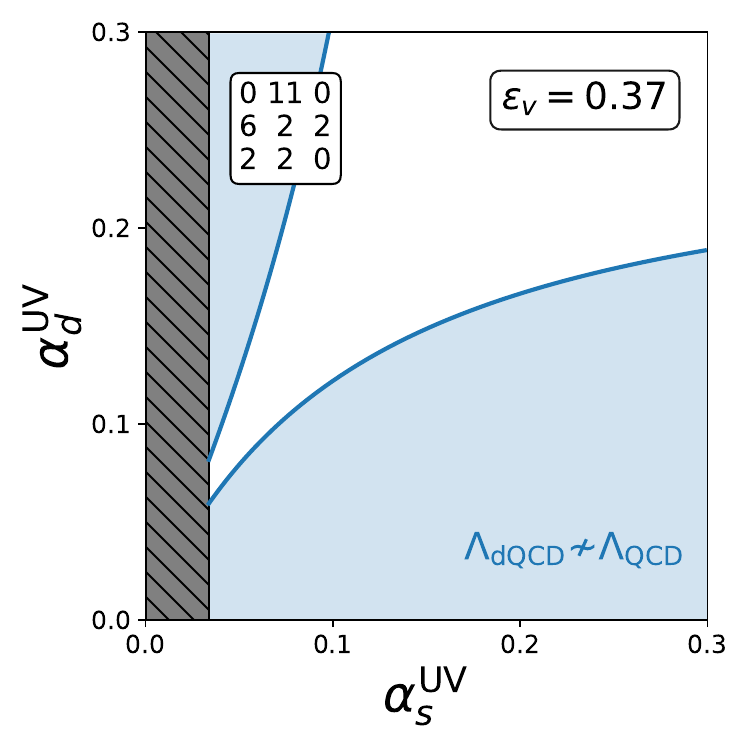}
    \caption{The UV coupling parameter space for a viable model that contains the field content required for our supersymmetric leptogenesis mechanism. The blue contours and grey hatched region are the same as in Fig.~\ref{fig:lowM_lowb_models}. The field content is specified as in Fig.~\ref{fig:SUSY_Nd4_max_epsv.pdf}.}
    \label{fig:LG_model_SUSY}
\end{figure}

We now consider embedding this set-up within a model with a decent value of $\epsilon_v$. We show such a model in Fig.~\ref{fig:LG_model_SUSY}. Its field content contains the two $\rep{3}{4}$ fields and two $\rep{3}{6}$ fields necessary for the above mechanism. In addition, it has two $\rep{6}{4}$ fields and two $\rep{6}{1}$ fields. We can arrange it so that these fields do not take part in the decay chains that redistribute the asymmetries, and so do not alter our earlier calculations: if their gauge quantum numbers are given by $\Phi_5 \sim \rep{6}{4}_{-2/3}$ and $\Phi_6 \sim \rep{6}{1}_{-2/3}$, and they interact through the superpotential terms
\begin{equation}
    W \supset \kappa_5\Phi_5\bar{\Phi}_4\bar{d} + \kappa_6\Phi_5\bar{\Phi}_6\bar{X} + \bar{\kappa}_6\bar{\Phi}_5\Phi_6X + \kappa_7\Phi_6\bar{d}\bar{d},
\end{equation}
then if $M_{\Phi_5} > M_{\Phi_4}, M_{\Phi_6}$, any initial thermal population of the new fields will decay into visible and dark baryons through the decays $\Phi_5 \rightarrow \Phi_4 + d$ and $\Phi_5 \rightarrow \Phi_6 + X$ followed by $\Phi_4 \rightarrow d + X$ and $\Phi_6 \rightarrow 2d$, where again we write the decays in terms of superfields as a shorthand. These additional interactions also respect the extended $\tilde{B}$ and $\tilde{L}$ symmetries, with $\tilde{B}(\Phi_5) = 5/12$, $\tilde{L}(\Phi_5) = 1/4$, $\tilde{B}(\Phi_6) = 2/3$, and $\tilde{L}(\Phi_6) = 0$.

\subsubsection{Cosmological constraints in supersymmetry}

We now discuss some of the additional challenges you face when implementing leptogenesis and asymmetric dark matter mechanisms in a supersymmetric theory: (i) the gravitino problem~\cite{Kawasaki:2008qe} and (ii) the relic density of the lightest supersymmetric particle (LSP). The latter issue applies to any supersymmetric ADM model, while the former is specific to our proposed leptogenesis-like mechanism; however, we mention them together as they can be solved simultaneously by considering a very light gravitino.

For the first of these problems, gravitinos in general can cause cosmological issues: if they are stable they are at danger of being overproduced, and the decays of unstable gravitinos can lead to the disruption of BBN. Avoiding these scenarios puts various upper bounds on the reheating temperature, with the latter case in particular requiring $T_\mathrm{RH} < 10^{6-7}$~GeV~\cite{Kawasaki:2008qe}. The gravitino problem is the conflict between this limit and the necessary RHN mass scale for successful thermal leptogenesis; this is especially problematic in our case, where for the $N_i$ to decay to the new heavy fields we need a large value of at least $T_\mathrm{RH} \gtrsim 10^{16}$~GeV. However, these limits are derived for a gravitino mass of 100~GeV---10~TeV, as is typical for gravity-mediated supersymmetry breaking. In gauge-mediated supersymmetry breaking scenarios the gravitino mass can be as low as a fraction of an eV, and a sufficiently light gravitino will decay prior to BBN and thus face no limit on the reheating temperature.

The second issue is that, for an ADM model to explain the cosmological coincidence, we need the dark matter density to consist predominantly of the dark baryons whose mass and number density are naturally related to those of the proton. However, in $R$-parity--conserving supersymmetry, the LSP is stable and its relic abundance also contributes to the cosmological dark matter density. We thus need to ensure that any stable LSP is only a subdominant component of the dark matter. A light gravitino with $m_{3/2} \lesssim 1$~keV decouples while relativistic, and so its relic density is estimated to be~\cite{Viel:2005qj}
\begin{equation}
    \Omega_{3/2}h^2 = \frac{m_{3/2}}{94\text{~eV}}\frac{10.75}{g_\star(T_D)},
\end{equation}
where $g_\star(T_D)$ is the number of relativistic degrees of freedom at its decoupling temperature. Taking $g_\star(T_D) \sim 90$~\cite{Ichikawa:2009ir}, the fraction of the DM relic density made up by the gravitino component is
\begin{equation}
    f_{3/2} \equiv \frac{\Omega_{3/2}}{\Omega_\mathrm{DM}} \sim \frac{m_{3/2}}{94\text{~eV}},
\end{equation}
where $\Omega_\mathrm{DM}h^2\sim0.12$~\cite{Planck:2018vyg}. Given its small mass, it will behave as a hot dark matter subcomponent, free-streaming and thus facing constraints from cosmic microwave background data and from large-scale structure. The most recent limit on its mass is $m_{3/2} < 2.3$~eV~\cite{Xu:2021rwg}, and so an eV-scale gravitino evades these bounds while making only a percent-level contribution to the present-day DM density and avoiding the gravitino problem.

Finally here we mention some alternative solutions to these problems. For the gravitino problem, a low reheating temperature is possible if the initial population of RHNs is produced non-thermally. This can occur naturally in supersymmetry by using oscillations along the flat directions of the superpotential~\cite{Giudice:2008gu}. To deal with the relic abundance of the LSP, we can consider models in which the LSP is unstable. This could occur if $R$-parity is broken, or if a different discrete symmetry such as baryon triality~\cite{Ibanez:1991hv} is imposed; however, in this case one would need to be careful to ensure the stability of the dark matter candidate.

\subsection{Annihilating the symmetric component of the dark matter}\label{sec:annihil_sym_DM}

In ADM models, we need to ensure that only the dark particle--anti-particle asymmetry contributes to the dark matter relic density, by requiring the symmetric component of the DM to annihilate away sufficiently quickly. As noted in Ref.~\cite{Bai:2013xga}, the dark baryons will annihilate efficiently into dark pions, which can then decay into SM fields via interactions mediated by the new jointly-charged fields. 

For the non-supersymmetric models where these new fields are TeV-scale, this decay can occur rapidly. In the model of Sec.~\ref{sec:lowM_LG}, the dark pion decay is mediated by the $\rep{3}{3}$ scalar $\Phi$ through the interaction $\kappa_3 \Phi \bar{d}_R X^c$. If the dark pion mass and decay constant are a factor of a few larger than their SM values, such that $m_{\pi_d} > 2m_s$, then the dark pion can decay into strange quarks with decay width $\Gamma \sim \kappa_3^4 f_{\pi_d}^2 m_{\pi_d} m_s^2/32\pi M_\Phi^4$. Taking $f_{\pi_d}, m_{\pi_d} = 300$~MeV, this gives a sufficiently short dark pion lifetime of
\begin{equation}
    \tau \approx \left(\frac{M_{\Phi}/\kappa}{\text{TeV}}\right)^4 \times 1 \times 10^{-6}\text{~s}.
\end{equation}

In the supersymmetric models we consider, however, the new jointly-charged fields are very heavy, and the dark pions are very long-lived. So, we must introduce another annihilation channel for the dark matter. One possibility that takes advantage of SUSY is to consider the nearly-minimal supersymmetric model, which introduces a gauge singlet field $S$ in addition to the MSSM particle content. In this theory the scalar component of the singlet is very light, and the dark baryons can decay into it via exchange of the fermionic singlinos~\cite{Haba:2010bm}.

Another general idea is to introduce a new dark gauge symmetry under which the dark baryons are charged, allowing them to annihilate to the dark force carriers. For this to efficiently eliminate the symmetric DM component, the annihilation cross-section must be at least a few times the weak scale~\cite{Graesser:2011wi}. For a gauge symmetry with a fine-structure constant $\alpha_X$, the cross-section $\sigma \sim \alpha_X/m_X^2$ will be sufficiently large for $\alpha_X \gtrsim 10^{-4}$. 

If this symmetry is broken, the massive gauge bosons can then decay to the SM. Considering the simplest case of a dark $U(1)_X$, the massive dark photon $Z'$ can decay to SM charged leptons via kinetic mixing with the photon. For a dark photon with mass $m_{Z'} \sim 250$~MeV, the decay width is $\Gamma \sim \alpha_\mathrm{EM} \epsilon^2 m_{Z'} / 3$, where $\epsilon$ is the kinetic mixing parameter. At this mass, a kinetic mixing parameter of $\epsilon \sim 10^{-8}$ evades current constraints~\cite{Fabbrichesi:2020wbt}, and also leads to a very short dark pion lifetime,
\begin{equation}
    \tau \approx \left(\frac{10^{-8}}{\epsilon}\right)^2\left(\frac{250\text{~MeV}}{m_{Z'}}\right) \times  1 \times 10^{-5}\text{~s}.
\end{equation}


\section{Phenomenology}\label{sec:pheno}

We now briefly discuss the phenomenology of the models in this framework. In this section we do not consider the non-supersymmetric models with TeV-scale masses for the new fields, as their experimental signatures have already been covered in Ref.~\cite{Bai:2013xga}. There, the relevant phenomenology arises from the new jointly-charged fields, which mediate DM-nucleon interactions in reach of current direct detection experiments, and can be produced directly at colliders.

In the supersymmetric models we consider, the new fields are very heavy, with a typical mass scale $M \sim 10^{16}$~GeV. This suppresses their interactions with SM particles, rendering them undetectable in current experiments. However, these models do still require new physics close to the TeV-scale that makes them accessible, both at colliders and in dark matter direct detection experiments. 

Firstly, these models exhibit the standard SUSY collider signatures. We have taken the scale of SUSY-breaking to be $M_\mathrm{SUSY} = 5$~TeV to lie above the current strongest constraints on the gluino mass~\cite{ATLAS:2022ihe}, but TeV-scale electroweakinos may be accessible in future experiments such as HL-LHC~\cite{GAMBIT:2023yih}.

Secondly, the efficient annihilation of the symmetric component of the dark matter to SM particles generally requires interactions mediated by new physics with masses at the TeV-scale or lower; these interactions also result in potentially detectable dark matter-nucleon scattering cross-sections. 

For example, in Section~\ref{sec:annihil_sym_DM} we introduced a broken dark gauge symmetry $U(1)_X$; in this case, the kinetic mixing between the photon and the massive dark photon $Z'$ allows the symmetric part of the DM to decay to SM charged leptons, and also mediates spin-independent DM-nucleon scattering. The scattering cross-section is given by $\sigma_{\chi N}^\mathrm{SI} \sim 16\pi\alpha_\mathrm{EM}\alpha_X\mu_{\chi N}^2\epsilon^2/m_{Z'}^4$, where $\chi$ is the dark baryon comprised of $N_d$ light dark quarks $X$, and $\mu_{\chi N}$ is the reduced mass of the DM and nucleon. Taking $m_\chi \sim 1$~GeV, and with an allowed kinetic mixing parameter of $\epsilon \sim 10^{-8}$ as before, we have
\begin{equation}
    \sigma_{\chi N} \approx \left(\frac{\alpha_X}{10^{-4}}\right)\left(\frac{\epsilon}{10^{-8}}\right)^2\left(\frac{100\text{~MeV}}{m_{Z'}}\right)^4 \times 1 \times 10^{-44} \text{~cm}^2.
\end{equation}
As before, the fine-structure constant $\alpha_X$ has a lower limit of $10^{-4}$ to ensure efficient annihilation of the symmetric SM component. With $\alpha_X \sim 10^{-4}$, for a 1~GeV dark matter particle the spin-independent scattering cross-section lies close to the neutrino fog~\cite{OHare:2021utq}, and below current experimental limits~\cite{Billard:2021uyg}. With a larger value of $\alpha_X$, or with a larger dark baryon mass\footnote{The required dark matter mass depends on the specific number density ratio $n_B/n_D$ resulting from a particular asymmetry generation mechanism. For an $\mathcal{O}(1)$ value of $n_B/n_D$, we can have $m_\chi$ up to $\sim$15~GeV.}, the dark matter could be in reach of future direct detection experiments.

All models in this framework also feature a potential stochastic gravitational wave (GW) signal arising from a first-order chiral phase transition in the dark sector \cite{Schwaller:2015tja, Breitbach:2018ddu, Croon:2018erz}. In the SM, it is known from lattice studies that the QCD phase transition is crossover \cite{Brown:1990ev, Stephanov:2006zvm}, and so does not generate such a signal. This is due to the specific, finite masses of the three light quark flavours, which control the dynamics of the transition. In the limit of massless quarks, a QCD-like group with 3 or more quark flavours will exhibit a first-order phase transition \cite{Pisarski:1983ms, Brown:1990ev}. So, the quark-hadron phase transition of the dark QCD $SU(N_d)$ gauge group will be first-order if $n_{dq} \geq 3$ and the light dark quarks are sufficiently light. Nearly all of the viable models we consider have sufficiently many light dark quarks: for example, looking at the viable supersymmetric models in Section~\ref{sec:SUSY_models}, for $N_d = 3$ only 3\% of the models with $\epsilon_v > 0.25$ have $n_{dq} < 3$, and for $N_d = 4$ all models have $n_{dq} \geq 6$.

During a cosmological first-order phase transition, gravitational waves are generated by the collisions of the expanding bubbles of broken phase, and by sound waves and turbulence in the surrounding plasma. For a particular dark sector, determining the strength of the stochastic GW signal requires the calculation of four parameters that characterise the dynamics of the phase transition: the energy released during the transition, $\alpha$; the terminal bubble wall velocity, $v_w$; the bubble nucleation temperature, $T_n$; and the duration of the phase transition, $\beta/H$. In Ref.~\cite{Rosenlyst:2023tyj}, calculations of these parameters were reviewed in the context of ADM theories like ours, in which the dark matter is a GeV-scale dark baryon and thus the dark QCD gauge group undergoes a GeV-scale chiral phase transition. There, it was claimed that $\mu$Hz GW signals could be generated that would be detectable at the proposed detector $\mu$Ares~\cite{Sesana:2019vho}.

However, this conclusion may be optimistic. Given that the dark QCD phase transition in our models has $T_n \sim \mathcal{O}$(GeV), the most important parameter in determining the strength and peak frequency of the GW signal is the phase transition duration $\beta/H$. The detectable $\mu$Hz GW spectra identified in Ref.~\cite{Rosenlyst:2023tyj} require $\beta/H \sim \mathcal{O}(10-100)$; however, studies of the phase transition in QCD-like theories based upon low energy effective models generally find a much shorter phase transition duration $\beta/H \sim \mathcal{O}(10^4)$. This is common to studies of color confinement in pure Yang-Mills gauge theories from $SU(3)$ \cite{He:2022amv, Morgante:2022zvc} up to $SU(8)$ \cite{Huang:2020crf}, and also to $SU(3)$ dark sectors with 3 or more light dark quarks \cite{Bai:2018dxf, Helmboldt:2019pan, Reichert:2021cvs, Pasechnik:2023hwv}. The GW spectrum generated by such a phase transition will have a higher peak frequency, typically in the mHz range where space-based interferometers like LISA \cite{Caprini:2019egz} and BBO \cite{Yagi:2011wg} are most sensitive. These GW signals generally lie below the sensitivity of the proposed detectors, but there are suggestions that they may be detectable at BBO with $T_n \sim \mathcal{O}$(10 GeV). We do note, finally, that these results are derived from effective field theories that attempt to capture complex non-perturbative behaviours. On the advent of more sophisticated, first-principles lattice calculations, it may be that these signals are stronger than expected and have greater prospects of detection.


\section{Conclusions}\label{sec:conclusions}

In this work, we explored a new dark QCD framework in which there is a dynamical mechanism to relate the confinement scales of visible and dark QCD. This provides a dark matter candidate, the dark baryon, whose mass is naturally on a similar order to the proton mass. When combined with an asymmetric dark matter mechanism that relates the number densities of visible and dark matter, this provides an explanation of the coincidental similarity of the cosmological mass densities of visible and dark matter. We followed on from previous work in which new field content is introduced such that the gauge couplings of visible and dark QCD evolve towards a fixed point with small, related values for the two couplings -- referred to as the finite-coupling fixed point approach -- and instead considered a framework in which the couplings at the infrared fixed point are both zero. In this approach, the initial coupling strengths in the UV both evolve to be very small, before the new field content decouples at a high mass scale. These couplings then evolve independently until the energy scale at which they become non-perturbative and confinement occurs. The visible and dark confinement scales can then be similar for a generic selection of initial values for the couplings in the UV, with the caveat that this depends strongly on the number of light dark quarks.

We analysed models in this framework in order to find viable models that could explain the similarity of the visible and dark confinement scales, looking at dark gauge groups with varying number of colours $N_d$, and adding in field content in various representations of $SU(3)_\mathrm{QCD} \times SU(N_d)_\mathrm{dQCD}$. We first looked at models where the required mass scale of new physics is $\mathcal{O}$(TeV) to avoid naturalness constraints from the running of the Higgs mass parameter. In that case we did find viable models, but they were rare, and required a close cancellation between the matter and gauge contributions to the $\beta$-function coefficients; this cancellation is similar to that required for the viable models in the FCFP approach. However, in the FCFP approach, the successful models required very large multiplicities for the new field content; here, that was no longer the case.

Secondly, we looked at models in which no such cancellation occurs between the various contributions to the $\beta$-function coefficients. In these models, the mass scale of new physics is generically very high (only a few orders of magnitude below the UV scale) and so to avoid the naturalness issue for the Higgs mass parameter required us to consider supersymmetric versions of these models. In this case, for different values of $N_d$ we considered generic sets of models with low multiplicities for the new field content we introduce. For each model we chose the number of light dark quarks to maximise the viability fraction $\epsilon_v$, our heuristic that characterises the ability of a given model to explain the similarity of the confinement scales. 

The results here were quite promising. In particular, when $N_d = 4$, we found that 65\% of models have $\epsilon_v > 0.3$, with most of these models requiring 11 light dark quarks. For $N_d = 3$ we found that 40\% of models have $\epsilon_v > 0.25$, with most models requiring 6 dark quarks. For $N_d = 2$, however, all but a few models were not viable.

These sets of models are intended to be amenable to further development into full models of asymmetric dark matter. In this direction, we showed examples of a leptogenesis-like asymmetry generation mechanism for viable models in both the non-SUSY and SUSY cases. We also highlighted some of the phenomenology of these models. In particular we considered the gravitational wave signals arising from the cosmological chiral phase transition in the dark QCD sector. In this framework this transition is generically first-order and thus features a stochastic GW signal. While these signals usually lie at mHz peak frequencies sensitive to space-based interferometers like LISA and BBO, they are thought to be too weak to be detectable, due to the short duration of the phase transition. However, depending on calculational uncertainties in the GW dynamics, there may be prospects for detection at BBO, or even the proposed $\mu$Ares detector in the $\mu$Hz region.

The proposed ADM mechanisms in this paper are not the only possible way to build upon models in this framework. Further studies of the phenomenology and cosmology of complete ADM models incorporating other ADM mechanisms would be welcome.

\section*{Acknowledgements}

ACR was supported by an Australian Government Research Training Program Scholarship. This research was supported by the Australian Government through the Australian Research Council Centre of Excellence for Dark Matter Particle Physics (CDM, CE200100008).

\bibliography{references}

\clearpage

\appendix

\onecolumngrid

\section{Models with sextet representations and decent values of $\epsilon_v^{10}$}\label{sec:sextet_model_appendix}

In Section~\ref{sec:lowM_models} we looked at non-SUSY models with $N_d = 3$, including new field content in at most sextet representations of visible and dark QCD. Allowing for at most three copies of each new field, and only specifying the field content charged under visible QCD, we find 242 models that have $b_v^{(0)}/2\pi = 1/12$. The hope is that these models may have a small enough one-loop $\beta$-function coefficient that they can have a decent proportion of viable parameter space with $M < 10$~TeV, $\epsilon_v^{10}$. When adding in dark field content to these models -- that is, both light and heavy $\rep{1}{3}$ fermions (dark quarks) and heavy $\rep{1}{6}$ fermions -- we find that only 16 of the models obtain a value of $\epsilon_v^{10} > 0.25$. In Table~\ref{tab:lowM_maxsextet_models} we show the field content of these 16 models, along with their values for $\epsilon_v$ and $\epsilon_v^{10}$.

\vspace{1cm}

\begin{table*}[h]
\begin{tabular}{ccc|cc}
    \textbf{fermions} & $n_{dq}$ & \textbf{scalars} & $\epsilon_v$ & $\epsilon_v^{10}$ \\ \hline
    $\begin{pmatrix}
       & 2 & 3\\
     6 & 0 & 0\\
     2 & 0 & 0
    \end{pmatrix}$
    & 1 &
    $\begin{pmatrix}
       & 0 & 0\\
     0 & 1 & 0\\
     0 & 0 & 0
    \end{pmatrix}$ & 0.42 & 0.28\\
    $\begin{pmatrix}
       & 2 & 3\\
     6 & 0 & 0\\
     2 & 0 & 0
    \end{pmatrix}$
    & 2 &
    $\begin{pmatrix}
       & 0 & 0\\
     3 & 0 & 0\\
     0 & 0 & 0
    \end{pmatrix}$ & 0.41 & 0.34\\
    $\begin{pmatrix}
       & 2 & 2\\
     6 & 1 & 0\\
     1 & 0 & 0
    \end{pmatrix}$
    & 1 &
    $\begin{pmatrix}
       & 0 & 0\\
     2 & 3 & 0\\
     0 & 0 & 0
    \end{pmatrix}$ & 0.37 & 0.28\\
    $\begin{pmatrix}
       & 2 & 1\\
     6 & 3 & 0\\
     0 & 0 & 0
    \end{pmatrix}$
    & 1 &
    $\begin{pmatrix}
       & 0 & 0\\
     1 & 2 & 0\\
     0 & 0 & 0
    \end{pmatrix}$ & 0.37 & 0.29\\
    $\begin{pmatrix}
       & 3 & 1\\
     6 & 3 & 0\\
     0 & 0 & 0
    \end{pmatrix}$
    & 3 &
    $\begin{pmatrix}
       & 0 & 0\\
     2 & 0 & 0\\
     1 & 0 & 0
    \end{pmatrix}$ & 0.38 & 0.29\\
    $\begin{pmatrix}
       & 3 & 2\\
     7 & 1 & 0\\
     1 & 0 & 0
    \end{pmatrix}$
    & 1 &
    $\begin{pmatrix}
       & 0 & 0\\
     1 & 2 & 0\\
     0 & 0 & 0
    \end{pmatrix}$ & 0.40 & 0.30\\
    $\begin{pmatrix}
       & 4 & 2\\
     7 & 1 & 0\\
     1 & 0 & 0
    \end{pmatrix}$
    & 3 &
    $\begin{pmatrix}
       & 0 & 0\\
     2 & 0 & 0\\
     1 & 0 & 0
    \end{pmatrix}$ & 0.40 & 0.29\\
    $\begin{pmatrix}
       & 3 & 1\\
     7 & 3 & 0\\
     0 & 0 & 0
    \end{pmatrix}$
    & 2 &
    $\begin{pmatrix}
       & 0 & 0\\
     3 & 0 & 0\\
     0 & 0 & 0
    \end{pmatrix}$ & 0.38 & 0.34\\ 
\end{tabular}
\quad
\begin{tabular}{ccc|cc}
    \textbf{fermions} & $n_{dq}$ & \textbf{scalars} & $\epsilon_v$ & $\epsilon_v^{10}$ \\ \hline
    $\begin{pmatrix}
       & 3 & 2\\
     8 & 1 & 0\\
     1 & 0 & 0
    \end{pmatrix}$
    & 3 &
    $\begin{pmatrix}
       & 0 & 0\\
     0 & 1 & 0\\
     0 & 0 & 0
    \end{pmatrix}$ & 0.39 & 0.33\\
    $\begin{pmatrix}
       & 4 & 2\\
     8 & 1 & 0\\
     1 & 0 & 0
    \end{pmatrix}$
    & 2 &
    $\begin{pmatrix}
       & 0 & 0\\
     3 & 0 & 0\\
     0 & 0 & 0
    \end{pmatrix}$ & 0.40 & 0.36\\
    $\begin{pmatrix}
       & 5 & 1\\
     8 & 2 & 0\\
     0 & 0 & 0
    \end{pmatrix}$
    & 1 &
    $\begin{pmatrix}
       & 0 & 0\\
     0 & 2 & 0\\
     1 & 0 & 0
    \end{pmatrix}$ & 0.37 & 0.27\\
    $\begin{pmatrix}
       & 4 & 1\\
     8 & 2 & 0\\
     0 & 0 & 0
    \end{pmatrix}$
    & 1 &
    $\begin{pmatrix}
       & 0 & 0\\
     2 & 3 & 0\\
     0 & 0 & 0
    \end{pmatrix}$ & 0.37 & 0.31\\
    $\begin{pmatrix}
       & 1 & 2\\
     8 & 2 & 0\\
     0 & 0 & 0
    \end{pmatrix}$
    & 1 &
    $\begin{pmatrix}
       & 0 & 0\\
     3 & 1 & 0\\
     1 & 0 & 0
    \end{pmatrix}$ & 0.39 & 0.27\\
    $\begin{pmatrix}
       & 2 & 3\\
     9 & 0 & 0\\
     1 & 0 & 0
    \end{pmatrix}$
    & 1 &
    $\begin{pmatrix}
       & 0 & 0\\
     3 & 1 & 0\\
     1 & 0 & 0
    \end{pmatrix}$ & 0.42 & 0.27\\
    $\begin{pmatrix}
       & 2 & 1\\
     9 & 2 & 0\\
     0 & 0 & 0
    \end{pmatrix}$
    & 1 &
    $\begin{pmatrix}
       & 0 & 0\\
     1 & 0 & 1\\
     0 & 0 & 0
    \end{pmatrix}$ & 0.37 & 0.27\\
    $\begin{pmatrix}
       & 4 & 1\\
     9 & 2 & 0\\
     0 & 0 & 0
    \end{pmatrix}$
    & 4 &
    $\begin{pmatrix}
       & 0 & 0\\
     1 & 2 & 0\\
     0 & 0 & 0
    \end{pmatrix}$ & 0.36 & 0.31\\
\end{tabular}
\caption{These sixteen models have $b_v^{(0)}/2\pi = 1/12\pi$ and at most three of each new field, where we only include fields in at most sextet representations of visible and dark QCD. For each model we have added dark field content to achieve a decent value for $\epsilon_v^{10}$, the viable proportion of UV coupling parameter space for which $M < 10$~TeV. For each model we give the viability fraction $\epsilon_v$ as well as $\epsilon_v^{10}$. The field content of each model is specified in the same way as in Table~\ref{tab:lowM_lowb_models}, but here the rows correspond to $\mathbf{R}_v = \mathbf{1}, \mathbf{3}, \mathbf{6}$ from top to bottom, and the columns correspond to $\mathbf{R}_d = \mathbf{1}, \mathbf{3}, \mathbf{6}$ from left to right.}
\label{tab:lowM_maxsextet_models}
\end{table*}

\end{document}